\documentclass[12pt,english]{extarticle}
\usepackage[T1]{fontenc}
\usepackage[latin9]{inputenc}
\usepackage{amsmath}
\usepackage{amssymb}
\usepackage{graphicx}
\usepackage{esint}

\makeatletter

\providecommand{\tabularnewline}{\\}

\newcommand{\lyxaddress}[1]{
\par {\raggedright #1
\vspace{1.4em}
\noindent\par}
}

\usepackage{babel}

\usepackage{lscape}

\makeatother

\begin{document}

\title{$c$-$a$-$ca$ Mean Field RVB Model of CuNCN Physics. Structure
Manifestations of the RVB Transitions.}

\maketitle
\today

\author{A.L. Tchougréeff$^{a,b}$ and R. Dronskowski$^{a}$}

\lyxaddress{$^{a}$Institut für anorganische Chemie RWTH-Aachen University, Landoltweg
1, D-52056, Aachen, Germany;\\
 $^{b}$Poncelet Lab., Independent University of Moscow, Moscow Center
for Continuous Mathematical Education, Moscow, Russia }
\begin{abstract}
We propose a new form of the frustrated Heisenberg antiferromagnetic
Hamiltonian with spatially anisotropic exchange parameters $J_{c},\, J_{a},\,\mathrm{and}\, J_{ac}$
extended along the $c$, $a$, and $a\pm c$ lattice directions and
apply it to describe fascinating physics of copper carbodiimide CuNCN
in the assumption of resonanting valence bond (RVB) type of its phases.
These are invoked to explain the intriguing absence of magnetic order
in CuNCN down to 4 K. We show that the quasiparticle spectrum of the
RVB model of the proposed Hamiltonian has three principal regimes:
(i) one with two pairs of lines of nodes, (ii) one with a pair of
lines of nodes (termed as 1D- and Q1D-RVB states), (iii) and one with
two pseudogaps and four nodal points (2D-RVB). We present a complete
parameters-temperature phase diagram of the $c$-$a$-$ca$-RVB model
constructed with use of the high-temperature expansion of the free
energy. The phase diagram thus obtained contains eight different phases
whose magnetic behavior includes Curie and Pauli paramagnetism (respectively,
in disordered and 1D- or Q1D-RVB phases), and gapped (quasi-Arrhenius)
paramagnetism (2D-RVB phases). Adding magnetostriction and elastic
terms to the free energy of the model we derive possible structural
manifestations of transitions between various RVB phases of the $c$-$a$-$ca$-model
of CuNCN. Assuming a sequence of transitions between RVB phases to
occur in CuNCN while temperature decreases explains the features observed
in the temperature runs of the magnetic susceptibility and lattice
constants. Confronting these with the magnetic susceptibility and
strucutre data measured as functions of temperature in the range between
$ca.$ 20 and 200 K we show a remarkably good agreement between our
theoretical predictions and the experiment as reached by ascribing
the model parameters values which are intuitively acceptable both
in terms of their absolute and relative magnitudes and of the character
of their geometry dependence.
\end{abstract}
\pagebreak{}
\tableofcontents{}
\pagebreak{}

\section{Introduction}

Recently CuNCN phase had been obtained and a series of measurements
had been performed of its spatial structure and magnetic susceptibility,
electric resistivity, heat capacity (all \emph{vs.} $T$) \cite{Dronskowski}.
Although on the basis of analogy with other materials of the \emph{M}NCN
series (\emph{M} = Mn, Fe, Co, Ni) one would expect more or less standard
antiferromagnetic behavior, it turned out that at low temperature
this material does not manifest any magnetic neutron scattering \cite{NoOrder}.
The plausible explanation of the latter fact as \textquotedbl{}absence
of local momenta\textquotedbl{} may be, however, misleading since
the absence of the magnetic scattering means only the absence of the
long range magnetic order (LRMO) (evanscence of the spin-spin correlation
function), not the absence of the local momenta themselves. Similar
situation can be observed in the crystals of Cu-carboxylate dimers
where local spins 1/2 do present on each Cu$^{2+}$ ion, but form
isolated singlet pairs so that no long range magnetic order does exist.
On the structural grounds one cannot expect anything like this in
CuNCN since among the contacts of individual Cu$^{2+}$ ions one cannot
select any which would be uniquely strong. This brought us\cite{Tch091,Tch096,Tch097,Tch099}
to the idea that the ground state of this material may be related
to the RVB state of the Cu$^{2+}$ local spins 1/2.

\begin{figure}
\center{\includegraphics[scale=0.6]{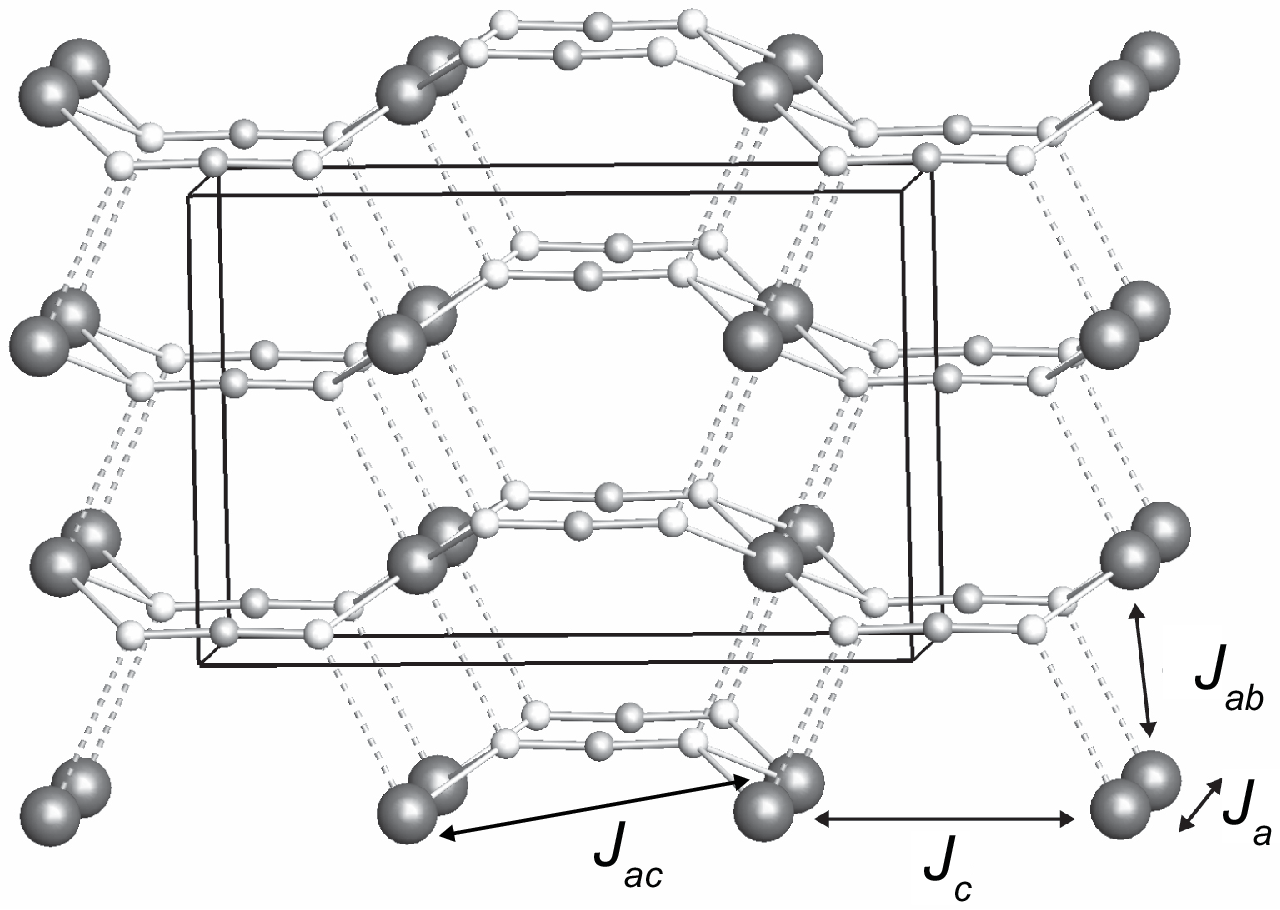}\\
 (a)\\
 \includegraphics[scale=0.5]{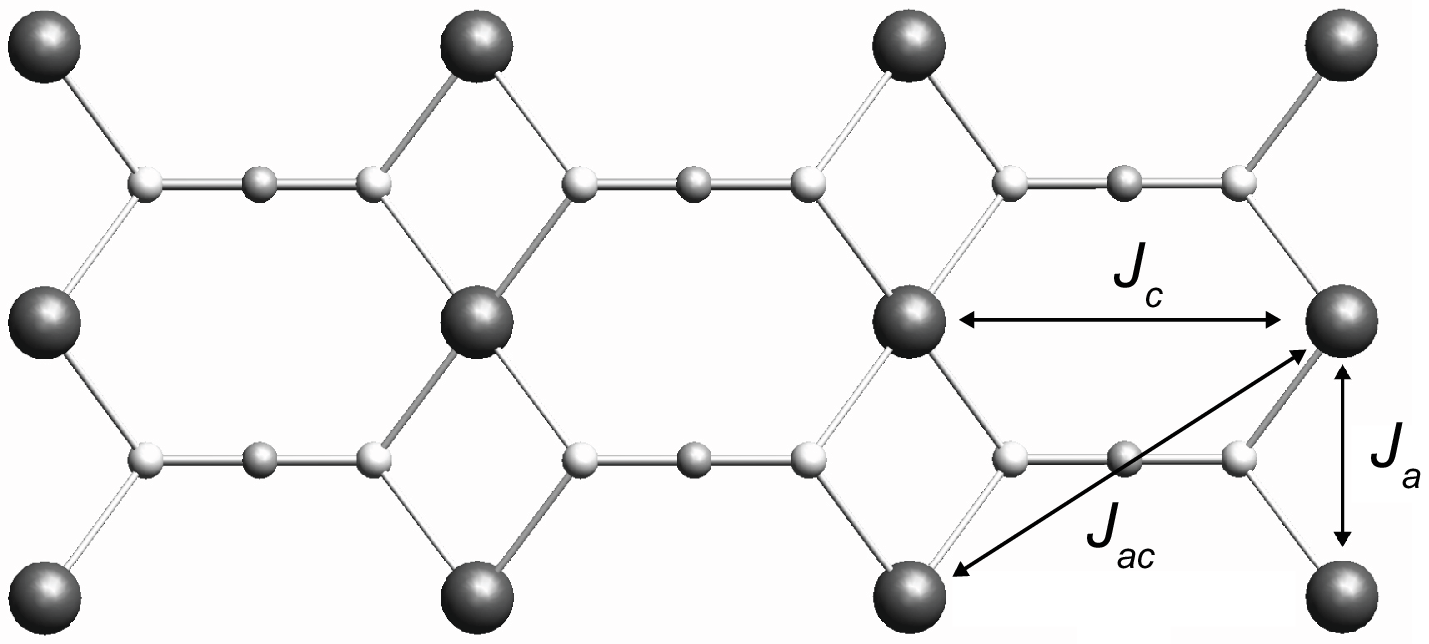} \\
 (b)\\
 \includegraphics[scale=0.5]{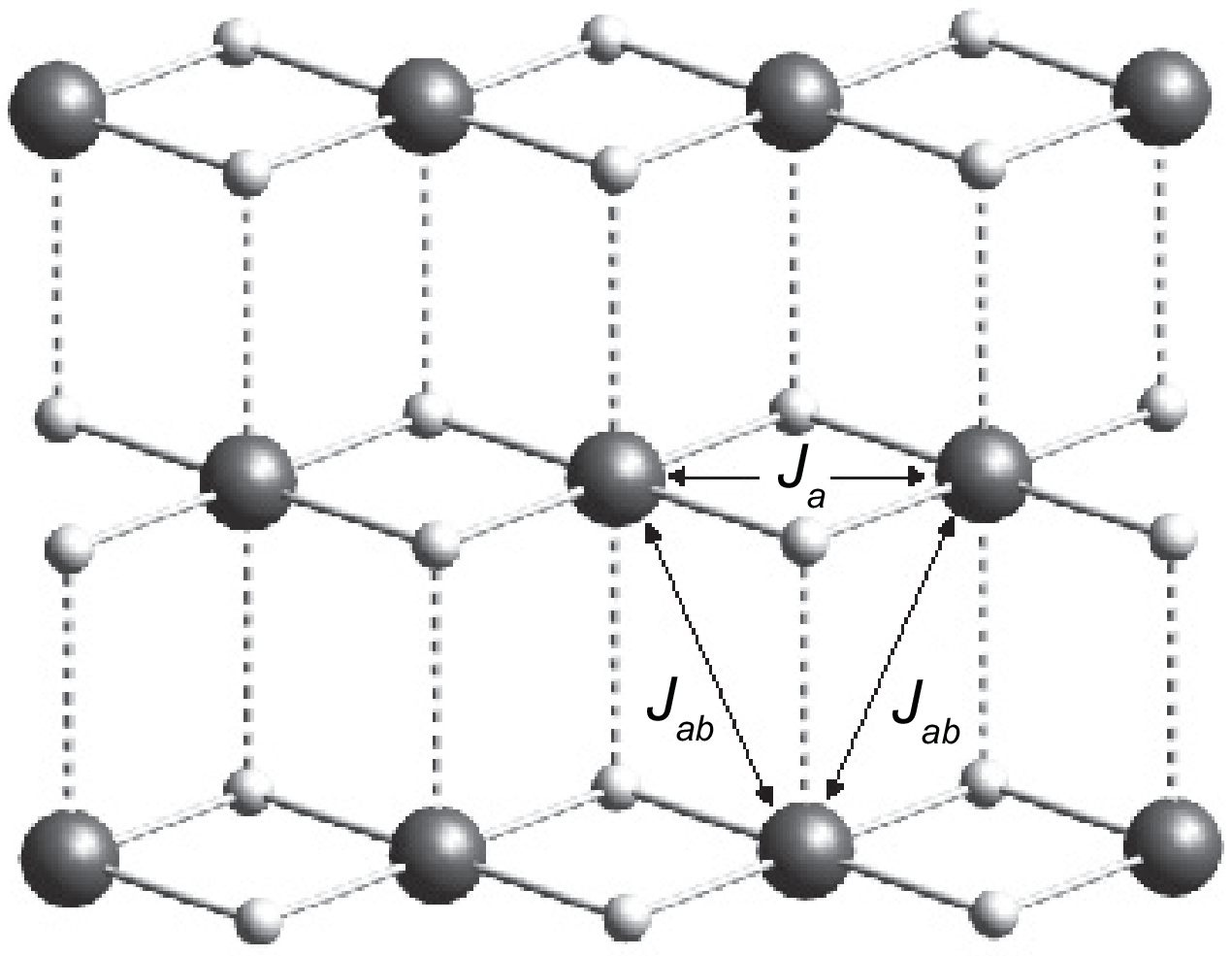}\\
 (c)}

\caption{The CuNCN crystal structure and the exchange parameters included in
the consideration. (a) shows the overall view on the structure. A
stronger $J_{c}$ extends in the $c$ direction; somewhat weaker $J_{a}$
extends in the $a$-direction. The weakest $J_{ab}$ extends along
the $b\pm a$ directions. (b) shows the $ac$ planes of CuNCN. The
interactions in all cases are mediated by the NCN$^{2-}$ moieties.
Two stronger interactions ($J_{c}$ and $J_{ac}$) are mediated by
the $\pi$-system of NCN$^{2-}$; somewhat weaker $J_{a}$ is strongly
contributed by a ferromagnetic cointerpoise terms dependent on the
hybridization at the N atoms. (c) shows for completeness the $ab$
planes of the CuNCN which are not considered in the present paper.}

\label{fig:CuNCN-ab-layer} 
\end{figure}

Previously we assumed \cite{Tch091,Tch096,Tch097,Tch099} that from
the materials structure Fig. \ref{fig:CuNCN-ab-layer} 
\cite{Dronskowski} one can conjecture an anisotropic triangular antiferromagnetic
Heisenberg model with the plane corresponding to the \emph{ab }plane
of the structure. This was not easy to reconsile with the intuitive
picture \cite{Goodenough} of the most important couplings to be extended
in the \emph{a} and \emph{c} directions. This brings us to the idea
to consider more antiferromagnetic couplings. This results in the
Heisenberg model with the Hamiltonian: 
\begin{equation}
\sum_{\mathbf{r}}\sum_{\mathbf{\tau}}J_{\mathbf{\boldsymbol{\tau}}}\mathbf{S}_{\mathbf{r}}\mathbf{S}_{\mathbf{r}+\mathbf{\mathbf{\tau}}}\label{eq:Hamiltonian}
\end{equation}
 where the translation vector $\mathbf{\tau}$ takes four values $\mathbf{\tau}_{i};i=1\div4;\mathbf{\tau}_{1}=(a,0);\mathbf{\tau}_{2}=(0,c);\mathbf{\tau}_{3}=(a,c);\mathbf{\tau}_{4}=(a,-c)$
with the interaction of the strength $J_{a}$ along the lattice vector
$\mathbf{\mathbf{\mathbf{\tau}}_{\mathrm{1}}}$ (two neighbors), with
a strength $J_{c}$ along the lattice vector $\mathbf{\mathbf{\mathbf{\tau}}_{\mathrm{2}}}$
(two neighbors as well), and interacion of the strength $J_{ac}$
along the lattice vectors $\mathbf{\mathbf{\mathbf{\tau}}_{\mathrm{3}}}$
and $\mathbf{\mathbf{\mathbf{\tau}}_{\mathrm{4}}}$ (two neighbors
along each). The importance of the diagonal $J_{ac}$ couplings in
the $\mathbf{\mathbf{\mathbf{\tau}}_{\mathrm{3}}}$ and $\mathbf{\mathbf{\mathbf{\tau}}_{\mathrm{4}}}$
directions have been recently reiterated in \cite{Tsirlin-arxiv}
although it is fairly in a line with the standard considerations of
\cite{Goodenough}. Either interactions along $\mathbf{\mathbf{\mathbf{\tau}}_{\mathrm{1}}}$
and $\mathbf{\mathbf{\mathbf{\tau}}_{\mathrm{2}}}$ or those along
$\mathbf{\mathbf{\mathbf{\tau}}_{\mathrm{3}}}$ and $\mathbf{\mathbf{\mathbf{\tau}}_{\mathrm{4}}}$
taken separately must lead to an antiferromagnetic state. However,
when considered simulateneously they interfere leading to a frustration
not allowing the spins to arrange in any LRMO state. For similar systems
a variety of RVB states have been proposed \cite{Ogata2003,Ogata&Fukuyama}.
Ground state of a similar, but spatially isotropic $J_{1}J_{2}J_{3}$
model have been treated recently by various methods and it has been
shown that spin-liquid states are very probable \cite{J1J2J3}. In
the present paper we consider in details the RVB states of the above
model in the RVB mean-field approximation and apply this to analysis
of the experimental data so far obtained for CuNCN.

\section{RVB mean-field analysis of the model }

\subsection{Quasiparticle spectrum}

Following the method \cite{Hayashi and Ogata } used by us previously
\cite{Tch091,Tch097,Tch099} we base the analysis of the Hamiltonian
eq. (\ref{eq:Hamiltonian}) on getting back to the fermion (spinon)
representation by the standard move: 
\begin{equation}
\mathbf{S}_{i}=\frac{1}{2}c_{i\alpha}^{+}\mathbf{\boldsymbol{\sigma}}_{\alpha\beta}c_{i\beta},\label{eq:SpinThroughFermi}
\end{equation}
 where $c_{i\sigma}^{+}(c_{i\sigma})$ are the fermion creation (annihilation)
operators; $\mathbf{\boldsymbol{\sigma}}_{\alpha\beta}$ are the elements
Pauli matrices and the summation over repeating indices is assumed.
Applying standard technique as described in Appendix \ref{sec:Equations-of-motion}
we reduce the problem to the set of $2\times2$ eigenvalue problems
for each wave vector $\boldsymbol{\mathbf{k}}$:

\[
\left(\begin{array}{cc}
\xi_{\mathbf{k}} & \Delta_{\mathbf{k}}\\
\Delta_{\mathbf{k}}^{*} & -\xi_{\mathbf{k}}
\end{array}\right)\left(\begin{array}{c}
u_{\mathbf{k}}\\
v_{\mathbf{k}}
\end{array}\right)=E_{\mathbf{k}}\left(\begin{array}{c}
u_{\mathbf{k}}\\
v_{\mathbf{k}}
\end{array}\right)
\]
 with 
\begin{eqnarray*}
\xi_{\mathbf{k}} & = & -3\sum_{\mathbf{\tau}}J_{\mathbf{\boldsymbol{\tau}}}\xi_{\mathbf{\boldsymbol{\tau}}}\cos(\mathbf{k\mathbf{\tau}})\\
\Delta_{\mathbf{k}} & = & 3\sum_{\mathbf{\tau}}J_{\mathbf{\boldsymbol{\tau}}}\Delta_{\mathbf{\boldsymbol{\tau}}}\cos(\mathbf{k\mathbf{\tau}})
\end{eqnarray*}
 (summation over $\mathbf{\boldsymbol{\tau}}$ extends to $\mathbf{\tau}_{i};i=1\div4$)
which results in the excitation spectrum of the form:

\begin{eqnarray*}
E_{\mathbf{k}} & =\pm & \sqrt{\xi_{\mathbf{k}}^{2}+\left|\Delta_{\mathbf{k}}\right|^{2}}
\end{eqnarray*}
whose eigenvectors are combinations of the destruction and creation
operators with the above Bogoliubov transformation coefficients. The
above set of equations closes by the selfconsistency conditions of
the form: 
\begin{eqnarray}
\xi_{\mathbf{\boldsymbol{\tau}}} & = & -\frac{1}{2N}\sum_{\mathbf{k}}\exp(i\mathbf{k\tau})\frac{\xi_{\mathbf{k}}}{E_{\mathbf{k}}}\tanh\left(\frac{E_{\mathbf{k}}}{2\theta}\right)\nonumber \\
\Delta_{\mathbf{\boldsymbol{\tau}}} & = & \frac{1}{2N}\sum_{\mathbf{k}}\exp(-i\mathbf{k\tau})\frac{\Delta_{\mathbf{k}}}{E_{\mathbf{k}}}\tanh\left(\frac{E_{\mathbf{k}}}{2\theta}\right)\label{eq:SelfConsistencyEquations}
\end{eqnarray}
for the order parameters (OPs) $\xi_{\tau},\Delta_{\tau}$. The lattice
symmetry considerations allow us to restrict ourselves by six OPs:
$\xi_{a},\xi_{c},\xi_{ac};\Delta_{a},\Delta_{c},\Delta_{ac}$. Using
the standard moves foreseen for the $SU(2)$ symmetric solutions as
described in Appendix \ref{sec:Equations-of-motion} we arrive to
the quasiparticle spectrum: 
\begin{eqnarray}
E_{\mathbf{k}}^{2} & = & 9\left(J_{a}^{2}\zeta_{a}^{2}\cos^{2}x+J_{c}^{2}\zeta_{c}^{2}\cos^{2}z+4J_{ac}^{2}\zeta_{ac}^{2}\cos^{2}x\cos^{2}z\right),\label{eq:Spectrum}
\end{eqnarray}
 where we set $x=\mathbf{k}_{x};z=\mathbf{k}_{z}$ and introduced
effective OPs $\zeta_{a}=\sqrt{\xi_{a}^{2}+\eta_{a}^{2}}$ , $\zeta_{c}=\sqrt{\xi_{c}^{2}+\eta_{c}^{2}}$,
and $\zeta_{ac}=\sqrt{\xi_{ac}^{2}+\eta_{ac}^{2}}$ .

\begin{landscape}
\begin{figure}
\begin{tabular}{|c|c|c|c|}
\hline 
\includegraphics[scale=0.5]{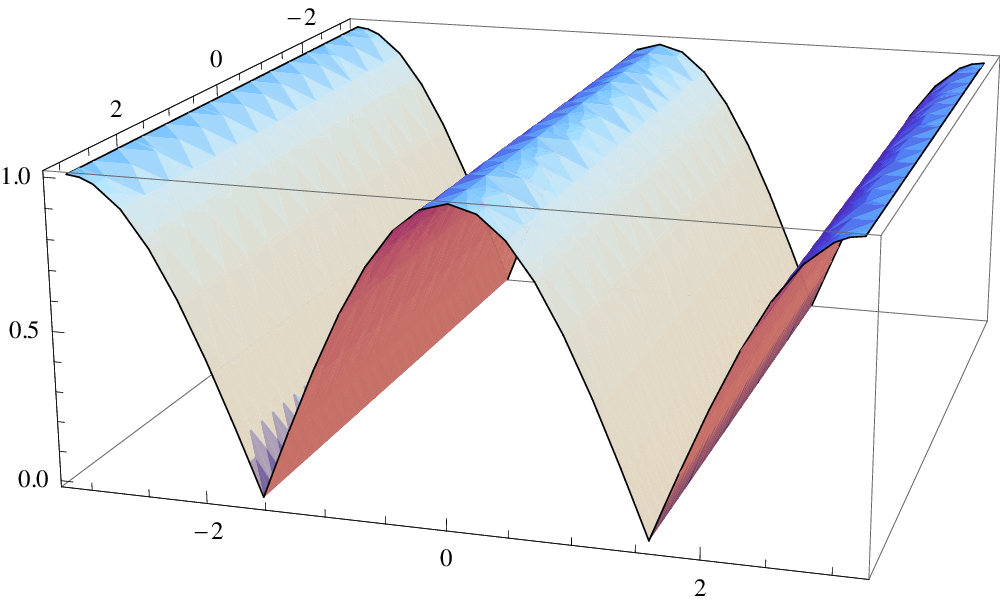} & \includegraphics[scale=0.5]{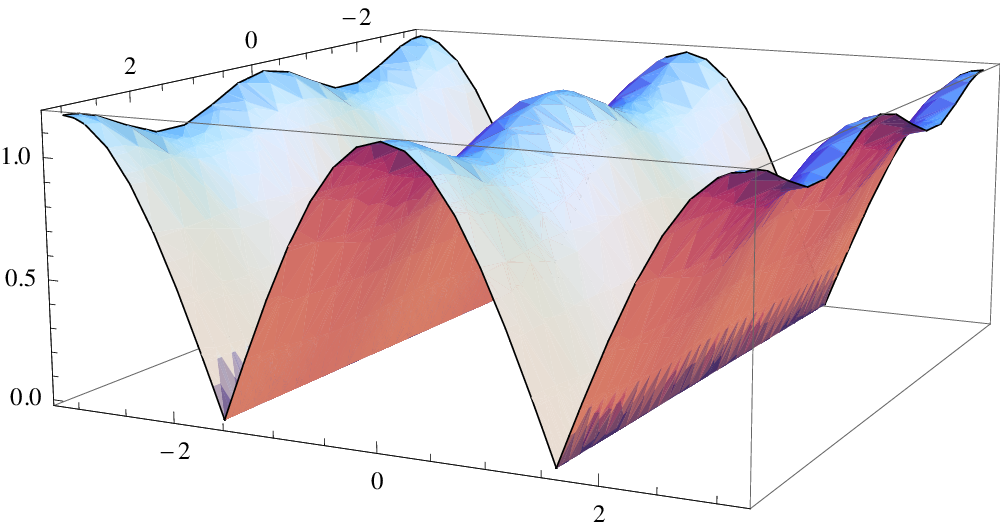} & \includegraphics[scale=0.5]{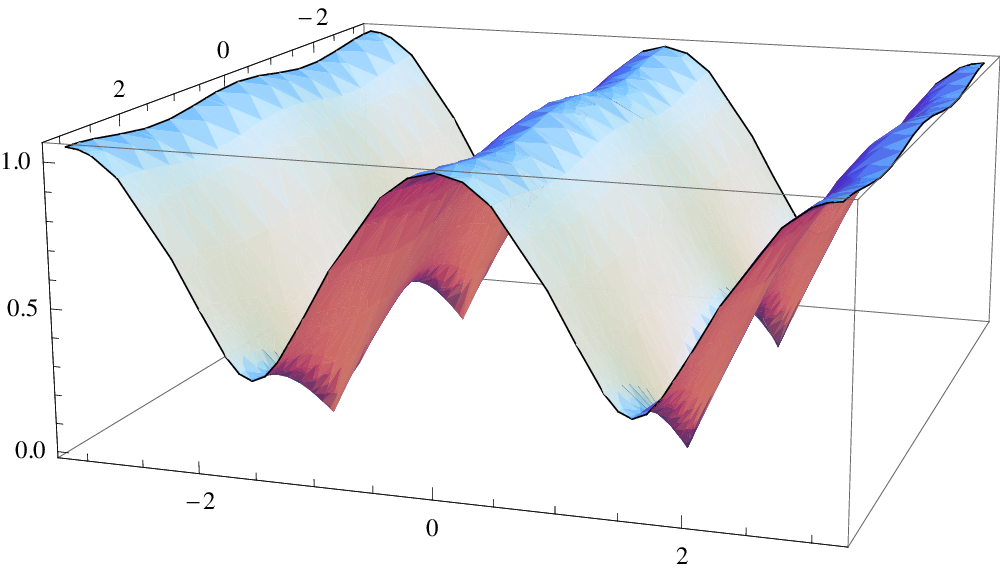} & \includegraphics[scale=0.5]{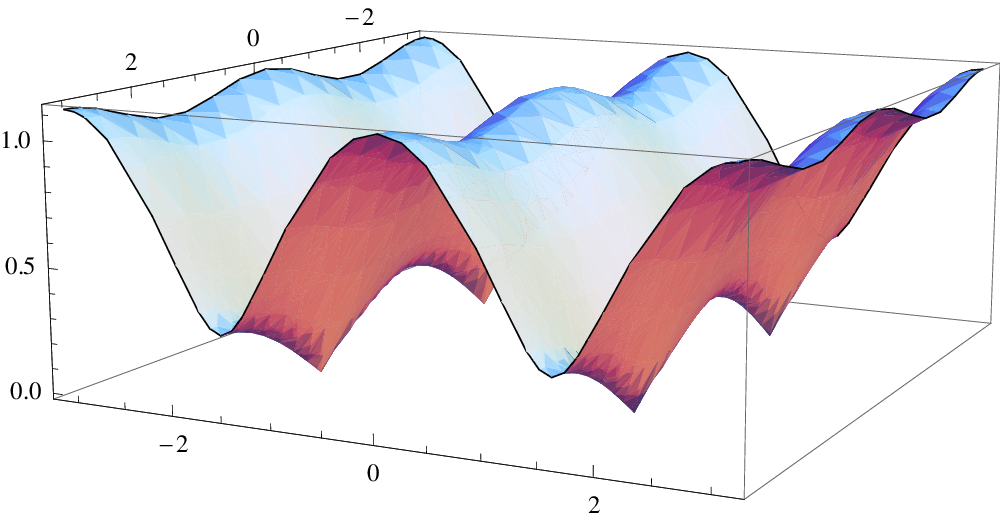}\tabularnewline
\hline 
\hline 
\includegraphics[scale=0.5]{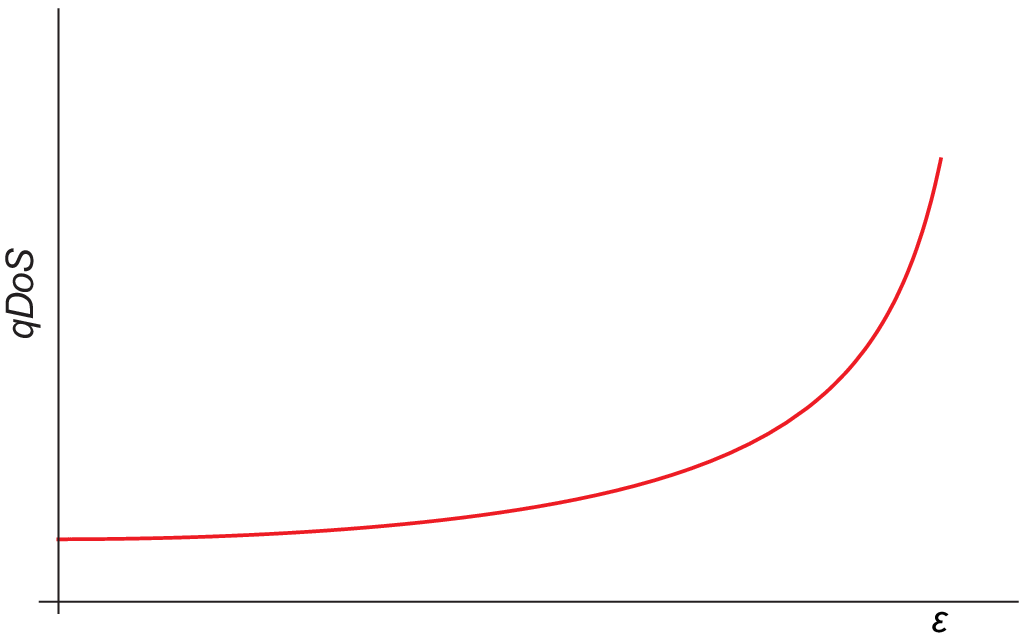} & \includegraphics[scale=0.5]{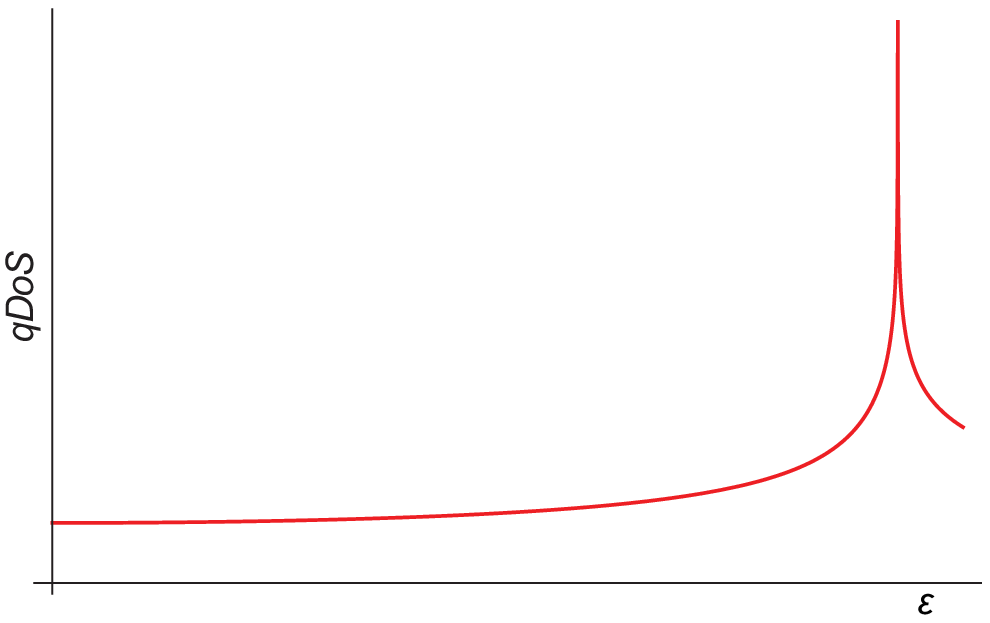} & \includegraphics[scale=0.5]{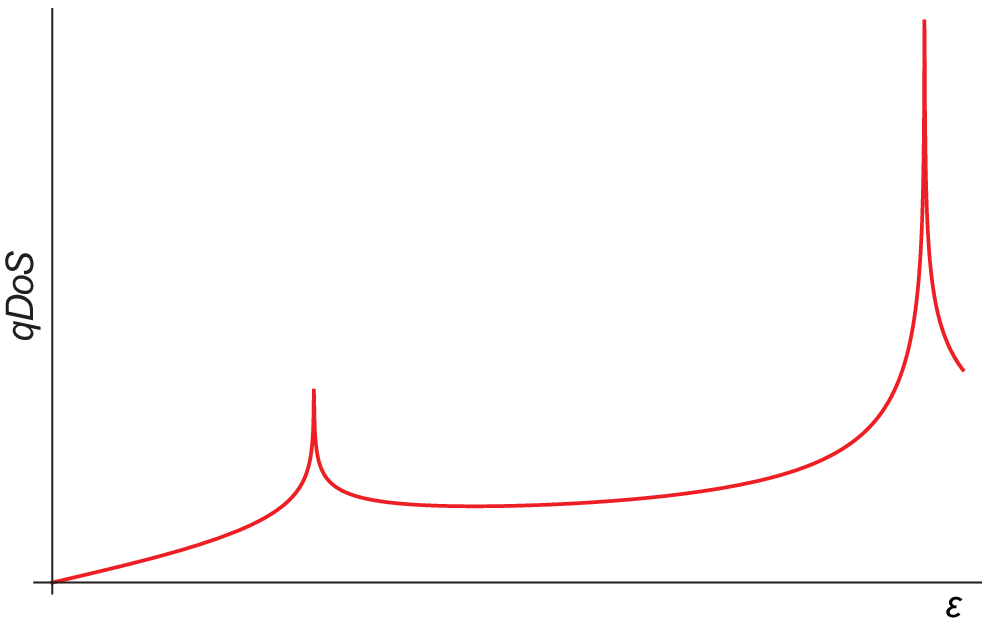} & \includegraphics[scale=0.5]{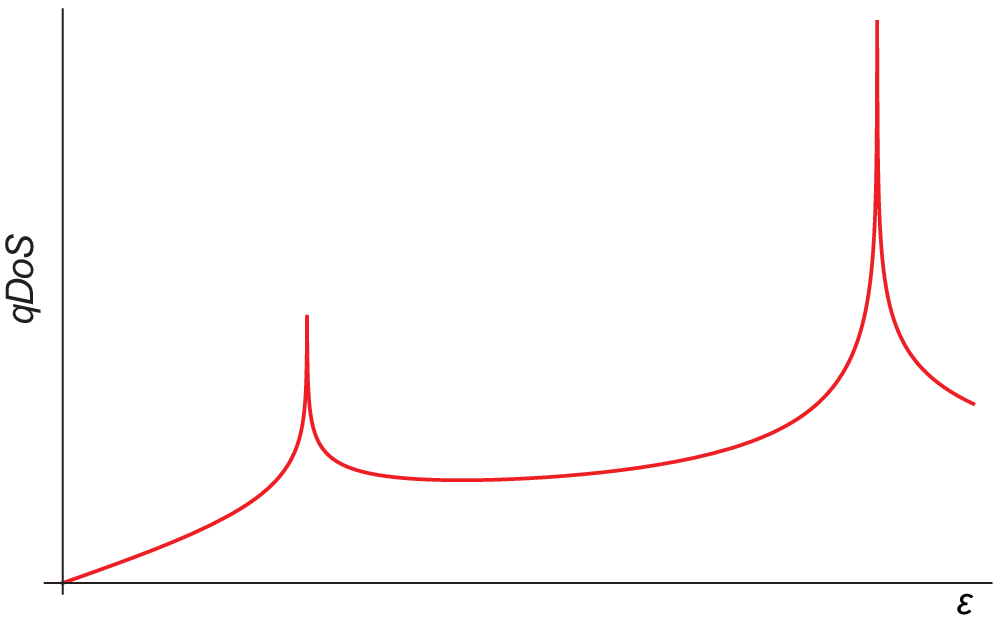}\tabularnewline
\hline 
1D-RVB $\zeta_{c}\neq0;\,\zeta_{a},\zeta_{ac}=0$ & Q1D-RVB $\zeta_{c},\zeta_{ac}\neq0;\,\zeta_{a}=0$ & 2D-RVB $\zeta_{c},\zeta_{a}\neq0;\,\zeta_{ac}=0$ & 2D-RVB $\zeta_{c},\zeta_{a},\zeta_{ac}\neq0$\tabularnewline
\hline 
\end{tabular}

\caption{Dispersion laws of the $c$-$a$-$ca$-RVB model for several exemplary
values of the pseudogap/bandwidth parameters $A=3J_{a}\zeta_{a};\, C=3J_{c}\zeta_{c};\, B=3J_{ac}\zeta_{ac}$
indicating characteristic feature of the quasiparticle spectrum in
different RVB states and the sketches of the relevant qDoS (see text
for the details). \label{fig:DispersionLaws}}
\end{figure}
\end{landscape}

The spectrum eq. (\ref{eq:Spectrum}) is depicted in Fig. \ref{fig:DispersionLaws}.If either of the OPs $\zeta_{a}$ or $\zeta_{c}$
is the only nonvanishing OP, the quasiparticle spectrum acquires corresponding
lines of nodes $z=\pm\frac{\pi}{2}$ or respectively $x=\pm\frac{\pi}{2}$
where the quasiparticles have zero energy. Since the dispersion of
quasiparticles takes place in only one crystallographic direction
(\emph{a} or \emph{c}) we spell these states as one-dimensional (1D-RVB)
states. In the low energy range the quasiparticle density of states
(qDoS) in the 1D-RVB states is constant. Due to the dispersionless
ridge in the spectrum the qDoS diverges at the ceiling of the quasiparticle
band \cite{Tch091}. This type of behavior is similar to that of the
spinons of the anisotropic triangular Heisenberg model as found in
\cite{Hayashi&Ogata-arxiv} and considered in our previous work \cite{Tch091,Tch096,Tch097,Tch099}
in relation to CuNCN. If both OPs $\zeta_{a,c}$ vanish and the OP
$\zeta_{ac}$ does not two paris of nodal lines exist along which
the quasiparticles have zero energy. In this satate the qDoS diverges
logarithmically at zero energy. We, however, do not focus on this
pecular state and it is not shown in Fig. \ref{fig:DispersionLaws}. 

If either of the nonvanishing OPs $\zeta_{a,c}$ is complemented by
the nonvanishing OP $\zeta_{ac}$ quasi-1D-RVB (Q1D-RVB) states appear.
The difference with the true 1D-RVB states is that in the Q1D-RVB
states there exists a nonvanishing dispersion in the direction transversal
to the node lines. At the higher energies the quasiparticle spectrum
of these states has maxima and saddle points instead of the ridge.
Thus the qDoS develops a finite hop at the ceiling of the quasiparticle
band and a van Hove singularty at somewhat lower energy, which serves
as a pseudogap. By contrast, if the nonvanishing OP's are $\zeta_{c}\,\mathrm{and}\,\zeta_{a}$
then irrespective to the value of the OP $\zeta_{ac}$ there are no
lines of nodes, but four nodal points ($\mathbf{k}=\left(\pm\frac{\pi}{2},\pm\frac{\pi}{2}\right)$)
of vanishing quasiparticle energies. In the vicinity of these points
the quasiparticles are massless as in the disordered (pseudometallic)
phase of the graphite monolayers \cite{Tch021}. The two possible
states of this type are spelled as 2D-RVB ones. The qDoS in 2D-RVB
states vanishes at the zero energy, being proportional to the energy
well below the smaller pseudogap. Otherwise the quasiparticle dispersion
law has saddle points at the two pseudogap energies and thus the qDoS
of 2D-RVB state develops two van Hove singularities at the corrsponding
pseudogaps.

\subsection{Free energy and phases of the model\label{sub:Free-energy-and}}

Following Ref. \cite{Ogata&Fukuyama} one can write immeditaly the
free energy of the \emph{c}-\emph{a}-\emph{ca} model in terms of the
above OPs: 
\begin{equation}
F=3J_{a}\zeta_{a}^{2}+3J_{c}\zeta_{c}^{2}+6J_{ac}\zeta_{ac}^{2}-\frac{2\theta}{4\pi^{2}}\intop_{BZ}\ln\left(2\cosh\left(\frac{E_{\mathbf{k}}}{2\theta}\right)\right)d^{2}\mathbf{k}.\label{eq:FreeEnergySimplified}
\end{equation}
where $\theta=k_{B}T$ (BZ stands for the integration over the Brillouin
zone). Minima of eq. (\ref{eq:FreeEnergySimplified}) with respect
to $\zeta$'s correspond to various possible phases of the system.
We postpone the study of the ground state (zero temperature limit)
of the present model to further publications and focus on the results
which can be obtained with use of the high temperature expansion (technicalities
are explained in Appendix \ref{sec:High-temperature-expansion.}). 

\begin{figure}
\includegraphics[scale=0.5]{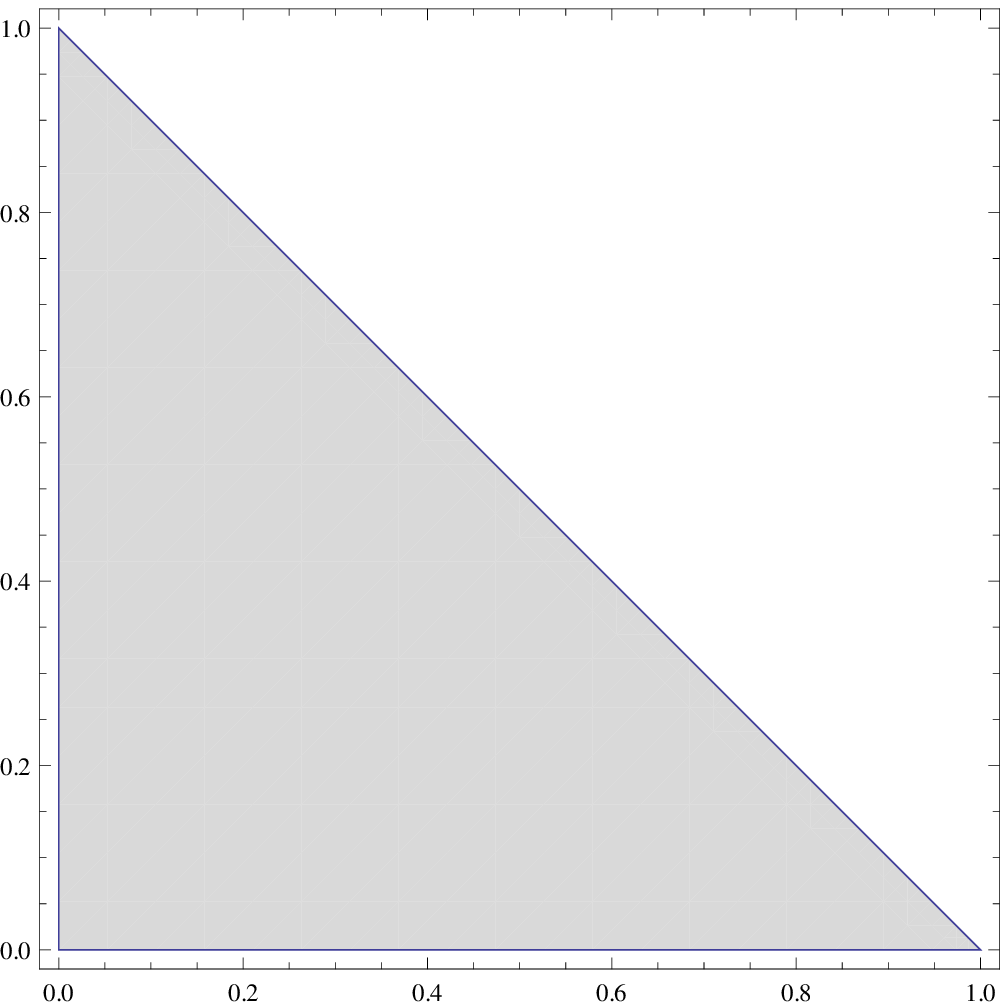}\includegraphics[scale=0.5]{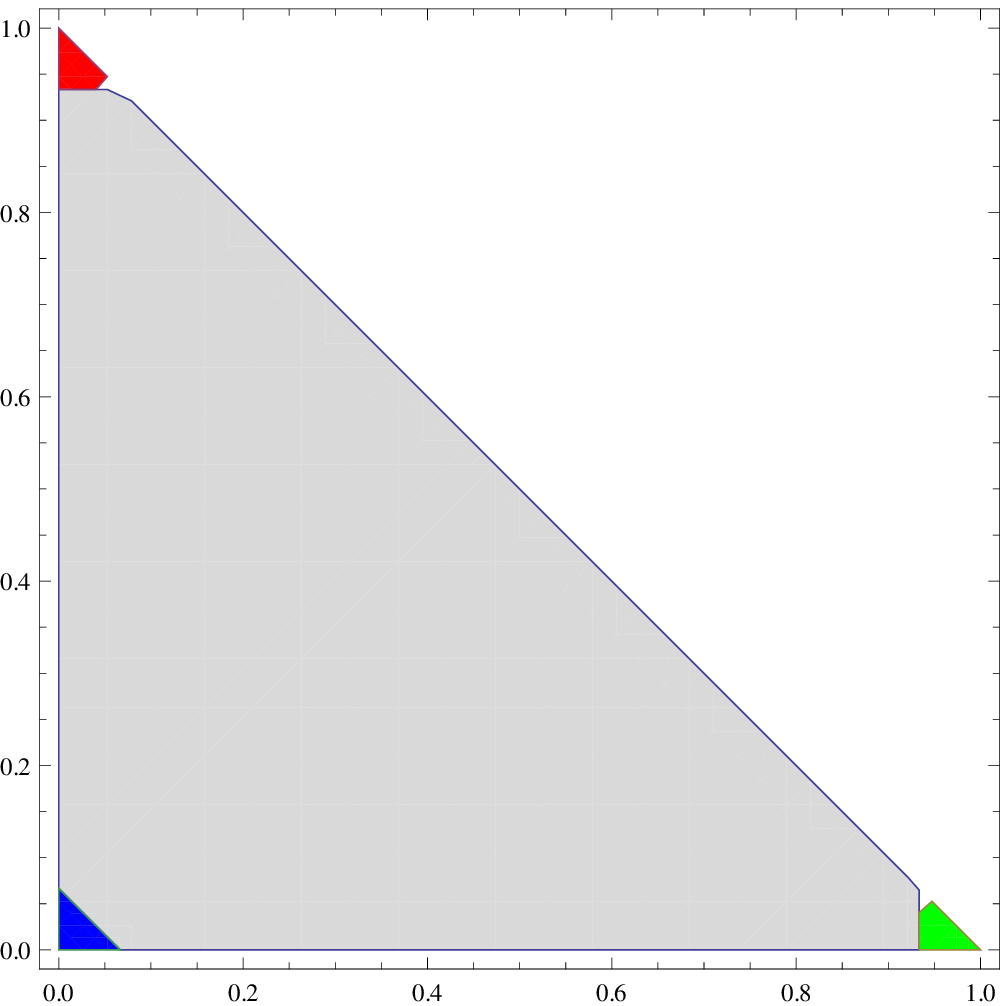}\includegraphics[scale=0.5]{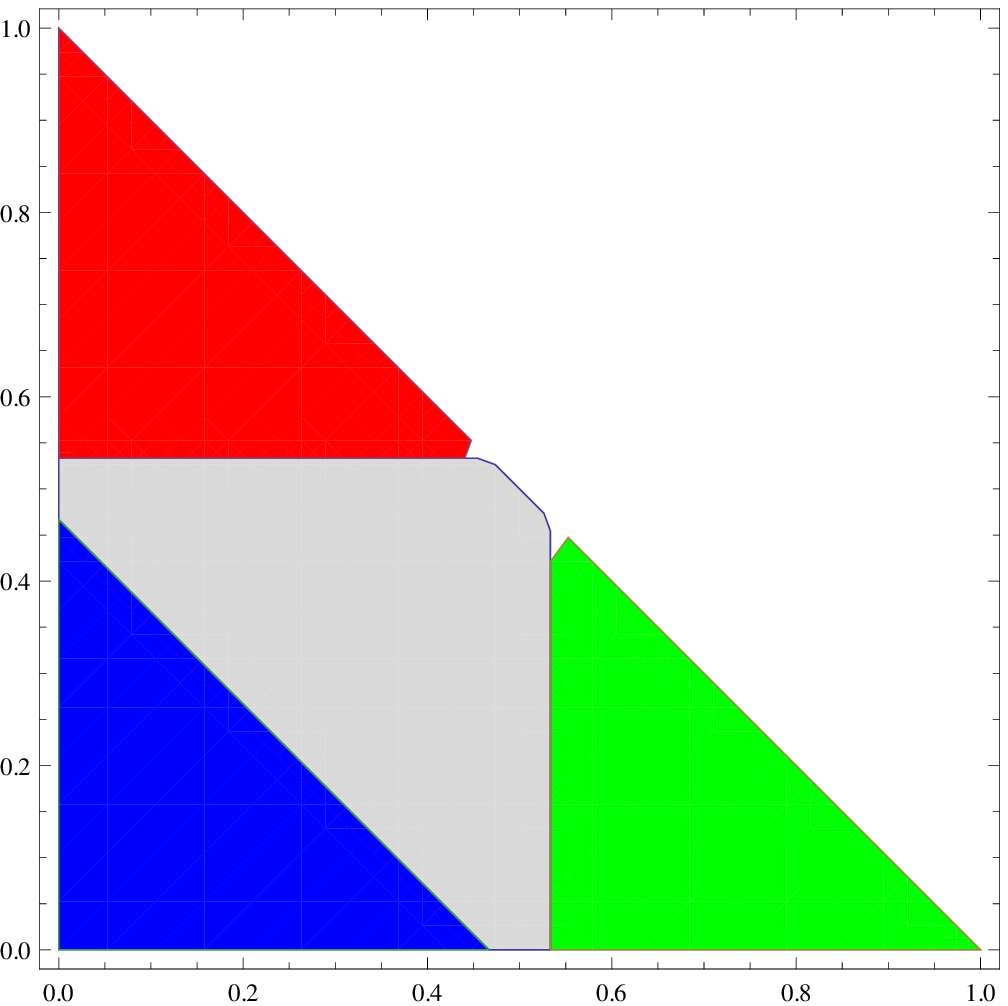}

\includegraphics[scale=0.5]{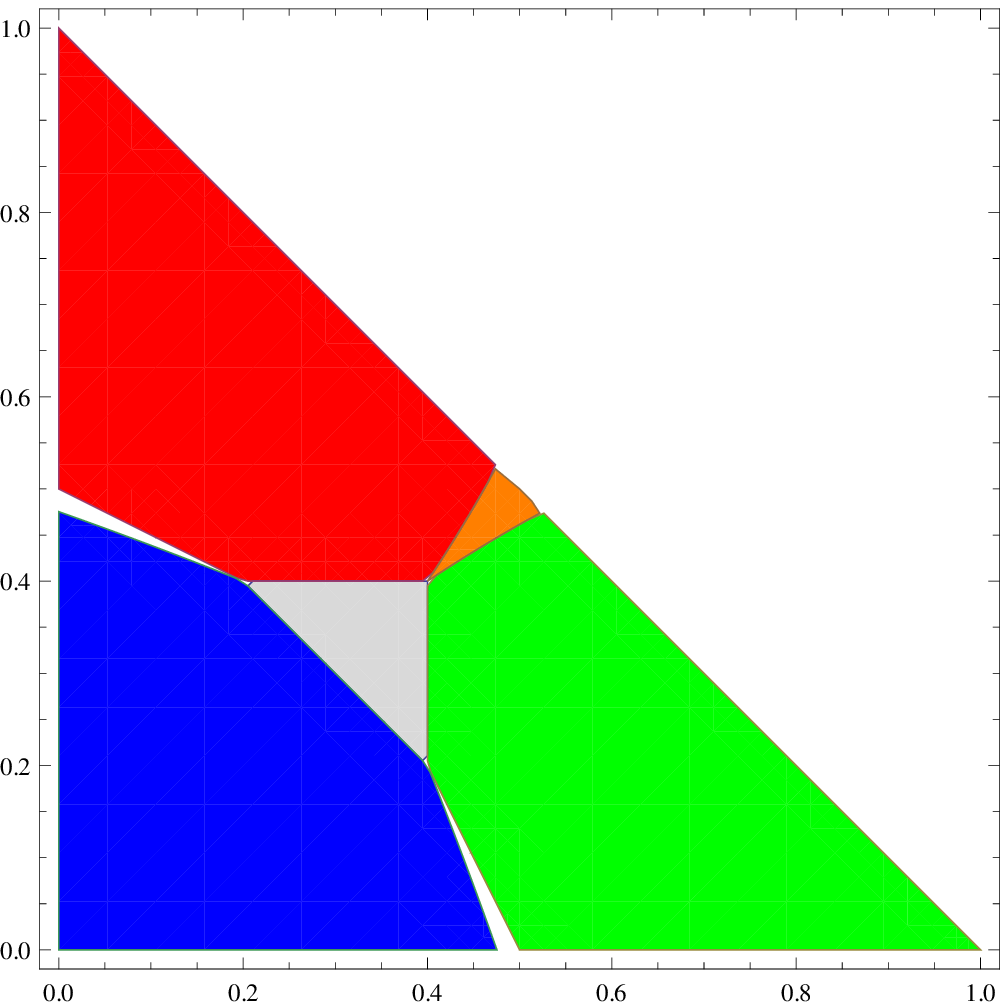}\includegraphics[scale=0.5]{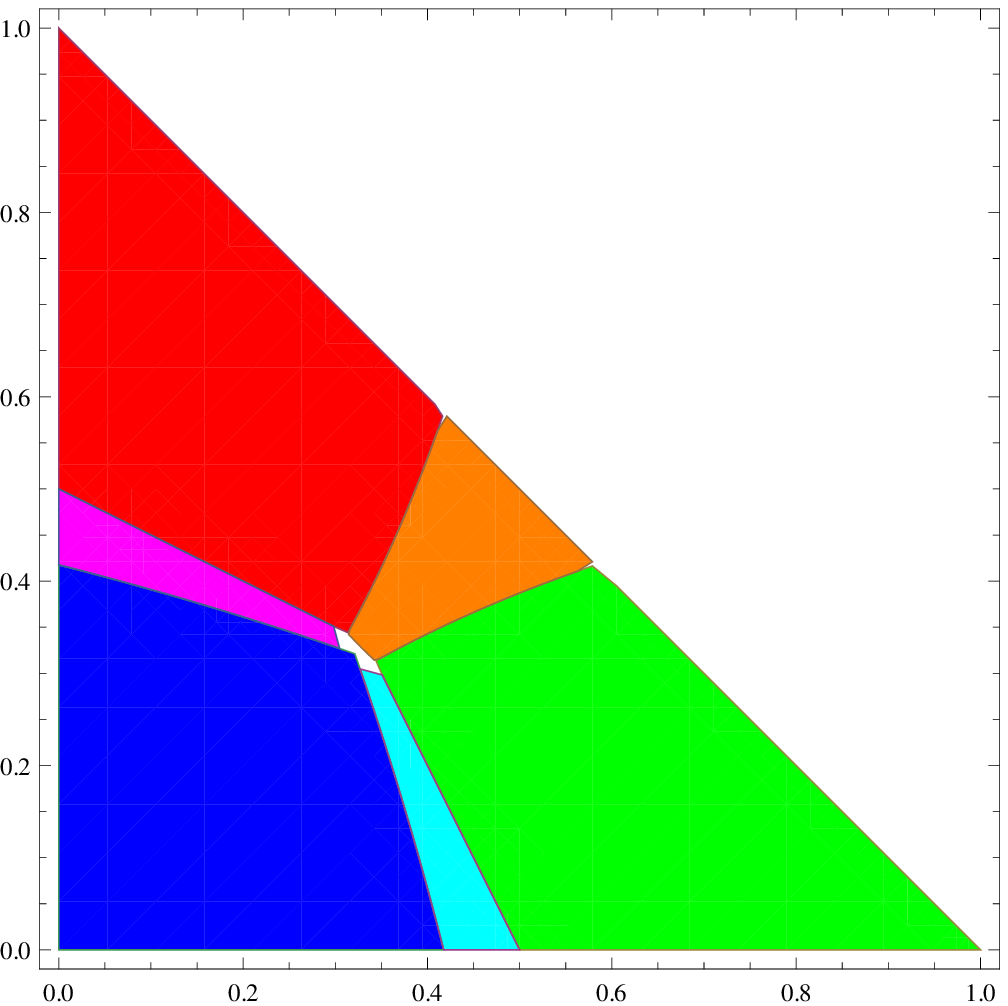}\includegraphics[scale=0.5]{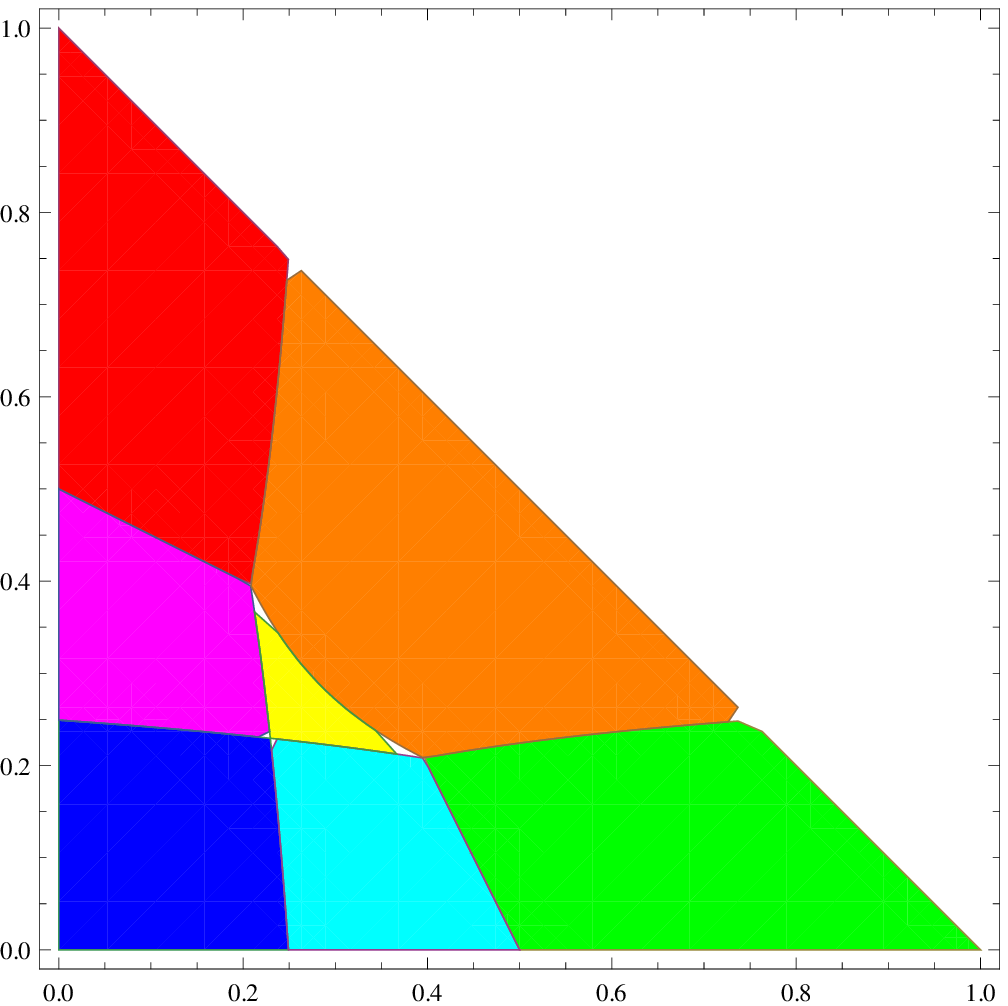}

\includegraphics[scale=0.5]{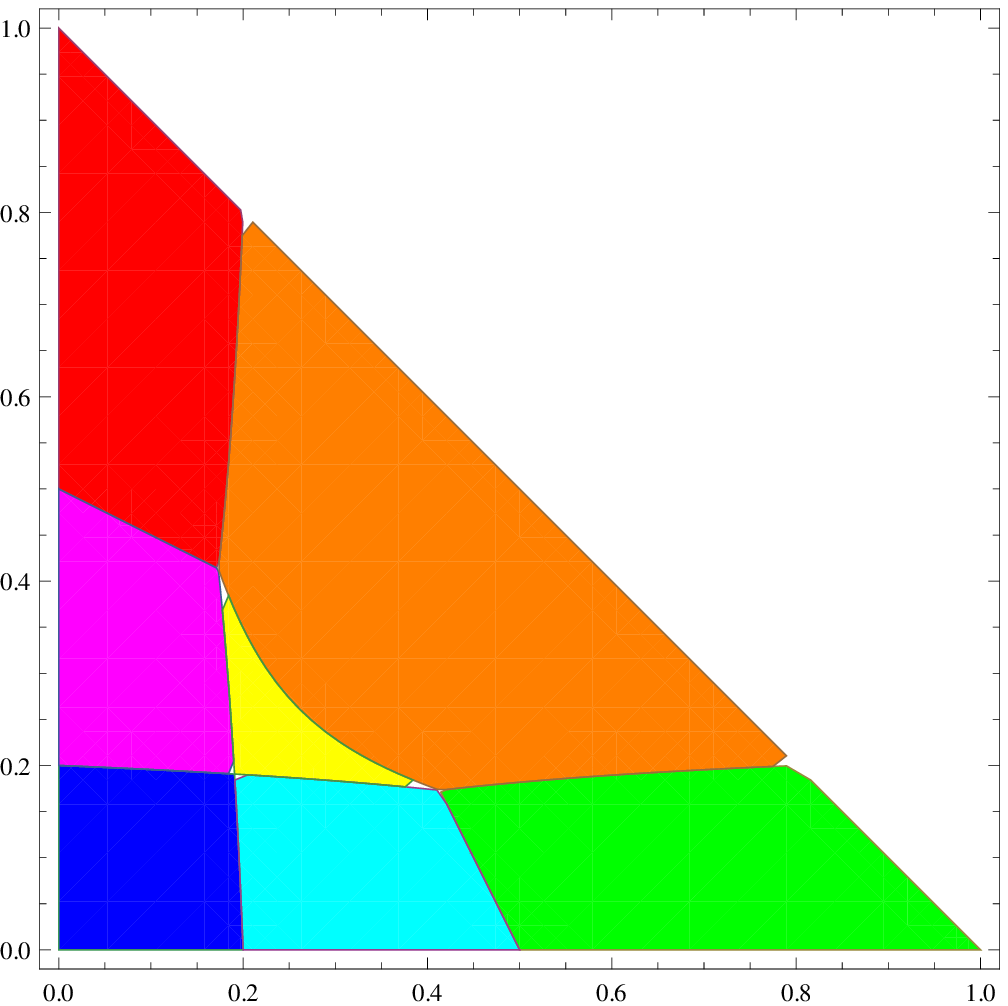}\includegraphics[scale=0.5]{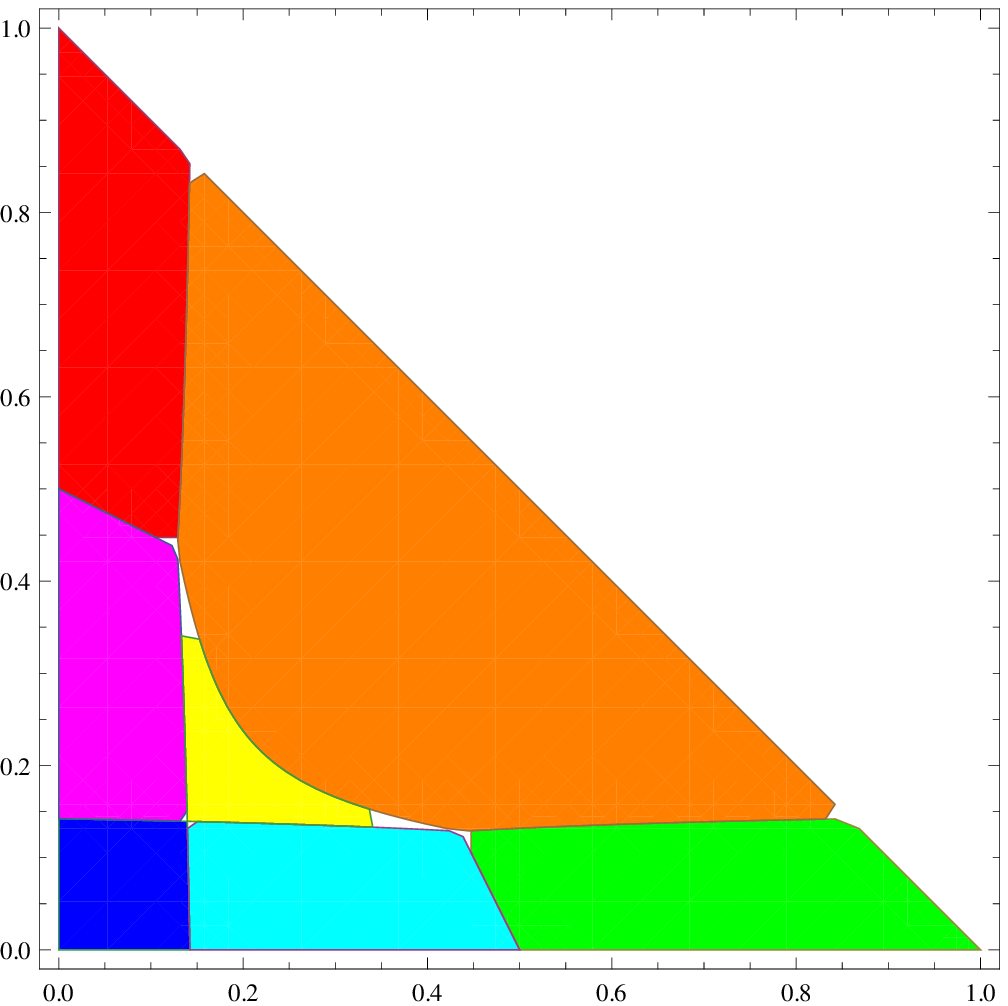}\includegraphics[scale=0.5]{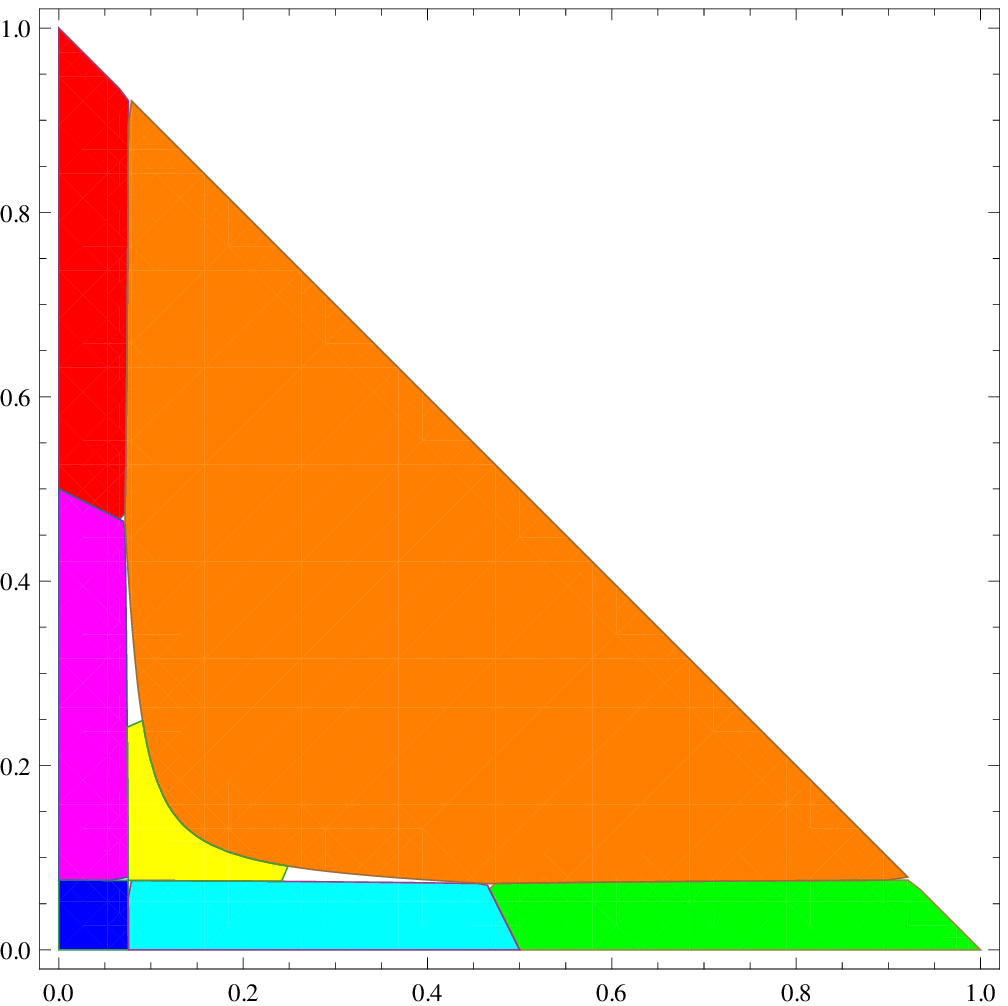}\caption{Parameter phase diagrams for the $c$-$a$-$ca$-RVB model in the
high-temperature approximation. The abscissa and ordinate represent
respectively $J_{a}$ and $J_{c}$; the entire parameters set is subject
to the condition $J_{a}+J_{c}+J_{ac}=1$. The temperatures as fractions
of $J_{a}+J_{c}+J_{ac}$ from the left upper diagram to the right
lower one are 0.4, 0.35, 0.2, 0.15, 0.1, 0.04, 0.03, 0.02, 0.01 (follow
rows). The colour coding for phases is: Curie paramagnetic, all OPs
are zero - gray, only one nonvanishing OP: red, green (1D-RVB, Pauli
paramagnet), blue for $\zeta_{c},\zeta_{a},\zeta_{ac}\neq0$, respectively;
one OP vanishing: magneta, cyan (Q1D-RVB, Pauli paramagnet), orange
for $\zeta_{a},\,\mathrm{or}\,\zeta_{c},\,\mathrm{or}\,\zeta_{ac}=0$,
respectively - observe the order of the list; yellow codes the phase
with three nonvanishing OPs. Orange and yellow phases (2D-RVB) feature
combination of the gapped Arrhenius-like temperature dependence of
magnetic susceptibility and its linear dependence well below the pseudogap. }

\label{Flo:ParameterPhaseDiagrams} 
\end{figure}

The results obtained by minimizing the high temperature expansion
of eq. (\ref{eq:FreeEnergySimplified}) with respect to the OPs are
depicted in the parameter phase diagram Fig. \ref{Flo:ParameterPhaseDiagrams}
and present in Table \ref{tab:OPsTemeperatureDependence}. Fig. \ref{Flo:ParameterPhaseDiagrams}
represents the phase diagram for the triple of exchange parameters
subject to the condition $J_{a}+J_{c}+J_{ac}=1$ and a series of temperatures
between 0.4 and 0.01 (the fractions of the above sum of the exchange
parameters is meant). One can see that eight phases are possible.
For the temperature above either of three critical ones:

\begin{equation}
\theta_{\mathbf{\tau}}^{\mathrm{crit}}=\frac{3}{8}J_{\mathbf{\tau}}.\label{eq:CriticalTemperaturesHighTemperature}
\end{equation}
the Curie paramagnetic phase: (1 - grey) in which all three OP's equal
to zero persists. 
Below these temperatures gapless phases appear in respective corners
of the parameter phase diagram. Their spectra are however quite different.
There are two 1D-RVB phases with OPs $\zeta_{c}\,\mathrm{or}\,\zeta_{a}\neq0$
(2 - red and 3 - green areas in Fig. \ref{Flo:ParameterPhaseDiagrams}).
Due to the constant qDoS at the zero energy they have to exhibit a
temperature independent paramagnetism which could be termed as Pauli
paramagnetism had it been due to conductivity electrons. Thus we spell
it as a quasi-Pauli one. The paramagnetic phase with $\zeta_{ac}\neq0$
(4 - blue area in Fig. \ref{Flo:ParameterPhaseDiagrams}) is quite
different: the maximum of the spectrum at the ceiling of the quasiparticle
band assures a constant value of the density of states (a finite hop)
rather a divergence as in 1D-RVB states, and logarithmically diverges
at the zero energy. Thus one has to expect the paramagnetic susceptibility
to logarithmically diverge at zero temperature in this phase. 

Below critical temperatures: 
\begin{equation}
\theta_{\mathbf{\tau\rightarrow\tau,\tau^{\prime}}}^{\mathrm{crit}}=\frac{J_{\mathbf{\tau}^{\prime}}}{8}\left(1-\frac{2J_{\mathbf{\tau}^{\prime}}}{3J_{\mathbf{\tau}}}\right)^{-1}=\frac{3J_{\mathbf{\tau}}J_{\mathbf{\tau}^{\prime}}}{8\left(3J_{\mathbf{\tau}}-2J_{\mathbf{\tau}^{\prime}}\right)}\label{eq:1OPto2OPCriticalTemperature}
\end{equation}
whose expression is already familiar from the high-temperature mean
field analysis \cite{Tch091,Tch096} of the RVB states in the triangular
anisotropic Heisenberg model, respective phases with two nonvanihsing
OP's $\zeta_{\tau}$ and $\zeta_{\tau^{\prime}}$ appear (the notation
refers to transition from the state where only one OP $\zeta_{\tau}\neq0$
to a state where two OP's $\zeta_{\tau},\zeta_{\tau^{\prime}}\neq0$).
The phases with two nonvanishing OPs as well are different. If one
of the nonvanishing OPs is $\zeta_{ac}$ (5 - magneta and 6 - cyan
areas in Fig. \ref{Flo:ParameterPhaseDiagrams}) the qDoS is constant
at the zero energy and thus quasi-Pauli paramagnetism has to be expected
in these Q1D-RVB phases. 

The phase with three nonvanishing OP's (8 - yellow) is a transient
one. It first appears below the \emph{octal} point $J_{a}=J_{ac}=J_{c}=\frac{1}{3};\theta^{*}=1/8$
where the Curie paramagnetic phase (grey) completely disappears and
shows up from the above Q1D-RVB phases (magenta or cyan) at the critical
temeperatures 

\begin{equation}
\theta_{\tau,ac\rightarrow a,c,ac}^{\mathrm{crit}}=\frac{3J_{ac}J_{\mathbf{\bar{\tau}}}}{8\left(3J_{ac}-2J_{\mathbf{\bar{\tau}}}\right)},\label{eq:2OPto3OPCriticalTemperature}
\end{equation}
($\bar{\tau}=c,a$ for $\tau=a,c$) but exists \emph{above} the critical
temperature of: 
\begin{equation}
\theta_{a,c,ac}^{\mathrm{crit}}=\frac{3J_{a}J_{ac}J_{c}}{8\left(3J_{a}J_{ac}-5J_{a}J_{c}+3J_{ac}J_{c}\right)}\label{eq:3OPsCriticalTemperature}
\end{equation}
where it switches to the 2D-RVB phase with only two nonvanishing OP's
$\zeta_{c}\,\mathrm{and}\,\zeta_{a}$ (7 - orange). That latter phase
appears also from the 1D-RVB phases (red and green) at the critical
temperatures given by eq. (\ref{eq:1OPto2OPCriticalTemperature}).
The only difference between the dispersion laws of these two phases
as depicted in 3-rd and 4-th columns of Fig. \ref{fig:DispersionLaws}
is somewhat more pronounced dispersion along a {}``ridge'' in the
case of the phase with three nonvanihsing OPs. Otherwise both 2D-RVB
phases have a qDoS with two van-Hove singularities at the energies
of their characteristic pseudogaps and the physics of these latter
two phases has to be pretty similar. It can be checked that the transitions
are largely of the second order, that is to say that the OPs split
from zero continuously at the corresponding transition temperatures.
The interphase borders between the 1D-RVB and Q1D-RVB phases (2/5
and 3/6 or red/magenta and gren/cyan, respectively) are special. Phases
2 and 3 are unstable towards developing the nonvanishing OP $\zeta_{ac}$
when the exchange parameter $J_{ac}$ turns to be larger than, respectively,
the exchange parameters $J_{c}$ or $J_{a}$ at whatever temperature.
On the lines $J_{ac}=J_{c}$ or $J_{ac}=J_{a}$ where the OP $\zeta_{ac}$
bounces from zero the transition temperatures from the Curie paramagnetic
state (phase 1 - grey) to either 1D-RVB or Q1D-RVB Pauli paramagnetic
phases pairwisely coincide (that is to 2 and 5 \emph{i.e.} to red
and magenta or to 3 and 6 \emph{i.e.} to gren and cyan). That means
that if the system parameters fall in the corresponding (magenta or
cyan) areas and it is cooled below the critical temperature eq. (\ref{eq:1OPto2OPCriticalTemperature})
with $\tau=ac,\tau^{\prime}=a,c$ , it directly goes from the Curie
paramagnetic phase to the corresponding Q1D-RVB phase, rather to a
1D-RVB phase.

In the last column of Table \ref{tab:OPsTemeperatureDependence} we
show the analytical forms of the qDoS characteristic for the respective
specific forms of the quasiparticle spectrum. It turned out quite
unexpectedly, that these qDoS can be found analytically for all phases
of the $c$-$a$-$ca$ RVB-model. Leaving the details of the derivation
for further publications we provide a sketch of the derivation in
Appendix \ref{sec:Quasiparticle-densities-of}.\begin{landscape}
\begin{table}
\caption{Temeperature dependencies of the OPs for possible phases of the $c$-$a$-$ca$-RVB
model in the high-temperature approximation. The areas of the existence
of the corresponding phases are those where the expressions under
the square roots are positive. The last column gives the qDoS in the
respective phases. There $\mathsf{K}$ stands for the complete elliptic
integral of the first kind. Details of derivation will be communicted
elsewhere. Their characteristic graphs are given in Fig. \ref{fig:DispersionLaws}.
\label{tab:OPsTemeperatureDependence} }

\begin{tabular}{|l|l|l|l|l|}
\hline 
 & No  & Color code  & OP's \emph{vs. $\theta$}  & $g(\varepsilon)$\tabularnewline
\hline 
\hline 
Curie  & 1  & Gray  & $\zeta_{c}=\zeta_{a}=\zeta_{ac}=0$  & $\delta(\varepsilon)$\tabularnewline
\hline 
Pauli  & 2  & Red  & $\zeta_{a}=\zeta_{ac}=0;\zeta_{c}=\frac{4\theta}{3J_{c}}\sqrt{1-8\theta/3J_{c}}$  & $\frac{2}{\pi}\frac{1}{\sqrt{C{}^{2}-\varepsilon^{2}}}$\tabularnewline
\hline 
Pauli  & 3  & Green  & $\zeta_{c}=\zeta_{ac}=0;\zeta_{a}=\frac{4\theta}{3J_{a}}\sqrt{1-8\theta/3J_{a}}$  & $\frac{2}{\pi}\frac{1}{\sqrt{A{}^{2}-\varepsilon^{2}}}$\tabularnewline
\hline 
$\frac{2\mu_{B}^{2}}{\pi^{2}B}\ln\frac{32B}{\pi e^{\gamma}\theta}$  & 4  & Blue  & $\zeta_{c}=\zeta_{a}=0;\zeta_{ac}=\frac{4\theta}{3J_{ac}}\sqrt{1-8\theta/3J_{ac}}$  & $\frac{2}{\pi^{2}B}\mathsf{K}\left(\frac{\sqrt{4B{}^{2}-\varepsilon^{2}}}{2B}\right)$\tabularnewline
\hline 
Pauli  & 5  & Magneta  & $\begin{cases}
\zeta_{a}=0\\
\zeta_{c}=\frac{4\theta\sqrt{3J_{ac}J_{c}-24\theta J_{ac}+16\theta J_{c}}}{3\sqrt{3}J_{c}\sqrt{J_{c}J_{ac}}}\\
\zeta_{ac}=\frac{8}{3}\sqrt{\frac{2}{3}}\frac{\theta}{J_{ac}}\sqrt{\frac{\theta\left(J_{ac}-J_{c}\right)}{J_{ac}J_{c}}}
\end{cases}$  & $\frac{4}{\pi^{2}}\frac{\mathsf{K}\left(\frac{2B}{\sqrt{4B{}^{2}+C{}^{2}-\varepsilon^{2}}}\right)}{\sqrt{4B{}^{2}+C{}^{2}-\varepsilon^{2}}}$\tabularnewline
\hline 
Pauli  & 6  & Cyan  & $\begin{cases}
\zeta_{a}=\frac{4\theta\sqrt{3J_{ac}J_{a}-24\theta J_{ac}+16\theta J_{a}}}{3\sqrt{3}J_{a}\sqrt{J_{a}J_{ac}}}\\
\zeta_{c}=0\\
\zeta_{ac}=\frac{8}{3}\sqrt{\frac{2}{3}}\frac{\theta}{J_{ac}}\sqrt{\frac{\theta\left(J_{ac}-J_{a}\right)}{J_{ac}J_{a}}}
\end{cases}$  & $\frac{4}{\pi^{2}}\frac{\mathsf{K}\left(\frac{2B}{\sqrt{4B{}^{2}+A{}^{2}-\varepsilon^{2}}}\right)}{\sqrt{4B{}^{2}+A{}^{2}-\varepsilon^{2}}}$\tabularnewline
\hline 
Arrhenius  & 7  & Orange  & $\begin{cases}
\zeta_{a}=\frac{4\theta\sqrt{3J_{a}J_{c}-24\theta J_{c}+16\theta J_{a}}}{3\sqrt{5}J_{a}\sqrt{J_{a}J_{c}}}\\
\zeta_{c}=\frac{4\theta\sqrt{3J_{a}J_{c}-24\theta J_{a}+16\theta J_{c}}}{3\sqrt{5}J_{c}\sqrt{J_{a}J_{c}}}\\
\zeta_{ac}=0
\end{cases}$  & $\begin{cases}
\frac{4\varepsilon}{\pi^{2}AC}\mathsf{K}\left(\frac{\varepsilon\sqrt{A^{2}+C^{2}-\varepsilon^{2}}}{AC}\right),\,\varepsilon<\min\left(A,C\right)\\
\frac{4}{\pi^{2}}\frac{\mathsf{K}\left(\frac{AC}{\varepsilon\sqrt{A^{2}+C^{2}-\varepsilon^{2}}}\right)}{\sqrt{A^{2}+C^{2}-\varepsilon^{2}}},\,\min\left(A,C\right)<\varepsilon<\max\left(A,C\right)\\
\frac{4\varepsilon}{\pi^{2}AC}\mathsf{K}\left(\frac{\varepsilon\sqrt{A^{2}+C^{2}-\varepsilon^{2}}}{AC}\right),\,\varepsilon>\max\left(A,C\right)
\end{cases}$\tabularnewline
\hline 
Arrhenius  & 8  & Yellow  & $\begin{cases}
\zeta_{a}=\frac{4\theta\sqrt{3J_{a}J_{ac}-24\theta J_{ac}+16\theta J_{a}}}{3\sqrt{3}J_{a}\sqrt{J_{a}J_{ac}}}\\
\zeta_{c}=\frac{4\theta\sqrt{3J_{c}J_{ac}-24\theta J_{ac}+16\theta J_{c}}}{3\sqrt{3}J_{c}\sqrt{J_{c}J_{ac}}}\\
\zeta_{ac}=\frac{4\theta\sqrt{24\theta J_{ac}J_{a}-40\theta J_{a}J_{c}+24\theta J_{ac}J_{c}-3J_{a}J_{ac}J_{c}}}{9J_{ac}\sqrt{J_{a}J_{c}J_{ac}}}
\end{cases}$  & $\begin{cases}
\frac{4}{\pi^{2}}\frac{\varepsilon\mathsf{K}\left(\sqrt{\frac{\varepsilon^{2}\left(A^{2}+4B^{2}+C^{2}-\varepsilon^{2}\right)}{A^{2}C^{2}+4B^{2}\varepsilon^{2}}}\right)}{\sqrt{A^{2}C^{2}+4B^{2}\varepsilon^{2}}},\,\varepsilon<\min\left(A,C\right)\\
\frac{4}{\pi^{2}}\frac{\mathsf{K}\left(\sqrt{\frac{A^{2}C^{2}+4B^{2}\varepsilon^{2}}{\varepsilon^{2}\left(A^{2}+4B^{2}+C^{2}-\varepsilon^{2}\right)}}\right)}{\sqrt{A^{2}+4B^{2}+C^{2}-\varepsilon^{2}}},\,\min\left(A,C\right)<\varepsilon<\max\left(A,C\right)\\
\frac{4}{\pi^{2}}\frac{\varepsilon\mathsf{K}\left(\sqrt{\frac{\varepsilon^{2}\left(A^{2}+4B^{2}+C^{2}-\varepsilon^{2}\right)}{A^{2}C^{2}+4B^{2}\varepsilon^{2}}}\right)}{\sqrt{A^{2}C^{2}+4B^{2}\varepsilon^{2}}},\,\varepsilon>\max\left(A,C\right)
\end{cases}$\tabularnewline
\hline 
\end{tabular}
\end{table}
\end{landscape}

\section{Physical properties within the model}

\subsection{Magnetic susceptibility}

We use the standard definition of the magnetic susceptibility per
spin: 
\begin{equation}
\chi=-2\mu_{B}^{2}\int g(\varepsilon)\frac{\partial f(\frac{\varepsilon}{\theta})}{\partial\varepsilon}d\varepsilon\label{eq:SusceptibilityDefinition}
\end{equation}
where $f(...)$ is the Fermi distribution function. The qualitative
behavior of the susceptibility as derived from the characteristic
features of the qDoS is designated in the first column of Table \ref{tab:OPsTemeperatureDependence}.
It is not surprizing that the temperature independent paramagnetism
(conditionally denoted as {}``Pauli'', although it goes about some
other quasipartiles rather the band electrons in metals) takes place
in the phases (Nos 2, 3, 5, and 6) with a constant density of states
at the zero energy. The gapless phase 4 with two pairs of intersecting
node lines manifests a logarithmic singularity in the qDoS at the
Fermi level. This singularity, however, integrates and produces a
logarithmic divergence of the susceptibility at the zero temperature.
For two phases (Nos 7 and 8) with nodal points in the dispersion law
and linear dependence of the qDoS in the low-energy range one has
to expect as well a linear dependence of the susceptibility on temperature
in the low-temperature region (well below the lower pseudogap) superimposed
with a quasi-Arrhenius bahavior with characteristic energy of the
pseudogap at higher temperature. That rich variety of possible phases
on the \emph{c-a-ca-}RVB model allows to eventually explain the\emph{
}magnetic behavior of CuNCN.

As we mentioned above and previously \cite{Tch091,Tch096,Tch097}
the absence of magnetic scattering in CuNCN is perfectly explained
by the hypothesis of the RVB character of its phases. The temperature
independent paramagnetism of CuNCN as observed at higher temperatures
is explained by formation of one of many 1D- or Q1D-RVB phases (Nos
2, 3, 5, and 6). That means that only one of the OPs $\zeta_{a}$
or $\zeta_{c}$ is nonvanishing. As previously we assume that this
phase sets on at some pretty high temperature which cannot be directly
checked due to decomposition of the material \cite{Tch096}. For a
pseudogap to open in a (Q)1D-RVB state with, say, nonvanishing $\zeta_{c}$
(phase 2, red or phase 5, magenta) one needs that at an observable
critical temperature the OP $\zeta_{a}$ splits from zero (phase 7,
orange or phase 8, yellow) since for the pseudogap to open it cannot
be the $\zeta_{ac}$ OP. Thus we assume the following form for the
temeperature dependence of the pseudogap $A$: 
\begin{equation}
A\left(\theta\right)=A_{0}\left(1-\frac{\theta}{\theta^{\mathrm{crit}}}\right)^{\nu}\label{eq:pseudogap-interpolation}
\end{equation}
below the critical temeperature $\theta^{\mathrm{crit}}$ (where by
$\theta^{\mathrm{crit}}$ one of the temperatures $\theta_{c\rightarrow c,a}^{\mathrm{crit}}$
or $\theta_{c,ac\rightarrow c,a,ac}^{\mathrm{crit}}$ is meant) and
perform the numerical integration of eq. (\ref{eq:SusceptibilityDefinition})
for the susceptibility with the qDoS for the RVB phase with two pseudogaps
where we also set $B=0$.%
\footnote{We notice that due to the character of the dependence of the OP $\zeta_{ac}$
on the model parameters shown in Table \ref{tab:OPsTemeperatureDependence}
and the plausible assumption of the relation $J_{ac}\gtrapprox J_{c}$
between the exchange parameter, which needs to hold for the Q1D-RVB
phase (magenta) to appear, this OP can never be large. Thus the characteristics
of the system are basically not affected by the specific value of
the $J_{ac}$ since its contribution is scaled down by the small value
of $\zeta_{ac}$. %
} 
\begin{figure}
\includegraphics{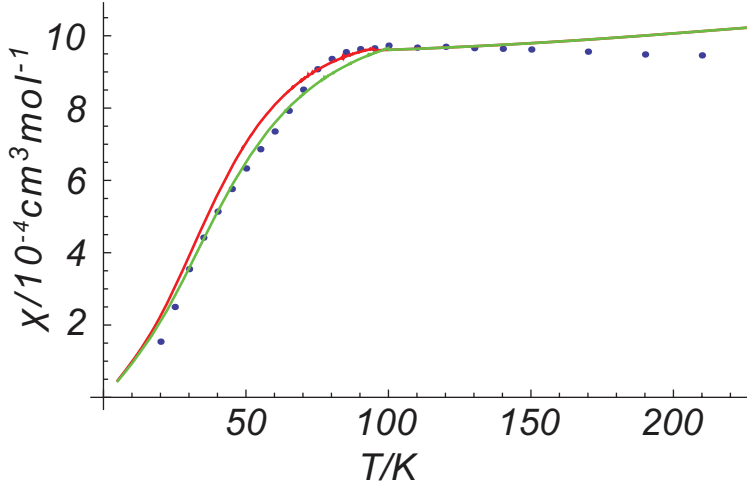}

\caption{EPR magnetic susceptibility of CuNCN Ref. \cite{Tch096} (blue dots)
as compared with the results of numerical integration with the qDoS
for the phase 7. One can easily see the linear tail in the low temperatue
range with no experimental points. Otherwise the parameters are $\theta^{\mathrm{crit}}=95$
K, $\nu=0.6$, $A_{0}=145$ K, $C=1250$ K (red line) and $\theta^{\mathrm{crit}}=100$
K, $\nu=0.5$, $A_{0}=150$ K, $C=1250$ K (green line). \label{fig:EPR-magnetic-susceptibility}}
\end{figure}

We performed several attempts and concluded that the values of $\theta^{\mathrm{crit}}$
($\theta_{c\rightarrow c,a}^{\mathrm{crit}}$ or $\theta_{c,ac\rightarrow c,a,ac}^{\mathrm{crit}}$),
$A_{0}$, $C$, and $\nu$ can be adjusted to reproduce the experimental
run of the susceptibility \cite{Tch096}. The results are shown in
Fig. \ref{fig:EPR-magnetic-susceptibility}. The value of $C$ ir
rather stable. Its scale is given by the magnitude of the Pauli paramagnetic
susceptiblity at a higer temperature and can be used to fit the parameters
of the original model eq. (\ref{eq:Hamiltonian}). The sets of parameters
$\theta^{\mathrm{crit}}=80$ K, $\nu=0.75$, $A_{0}=240$ K or $\theta^{\mathrm{crit}}=100$
K, $\nu=0.5$, $A_{0}=150$ K, as well as $\nu=0.7$, $A_{0}=170$
K; $\nu=0.6$, $A_{0}=145$ K are equally good in terms of describing
the susceptibility. The fact that the classical exponent $\frac{1}{2}$
coming from our simple high-temperature treatment allows for an acceptable
fit of experimental susceptibility is pretty remarkable. We assume
at this point for the sake of simplicity that the phase with the finite
qDoS at the zero energy is the 1D-RVB one with the nonvanishing $\zeta_{c}$.
Then using the zero temperature limiting value of this OP $\zeta_{c}=\frac{1}{\pi}$
\cite{Tch091} we get $J_{c}=1310$ K. This nicely agrees with the
original estimate of \cite{Dronskowski} (\emph{ca.} 1000 K) and is
more or less supported by other sources \cite{Tsirlin-arxiv}. That
is also what one can intuitively expect relying on the Goodenough-Kanamori
rules \cite{Goodenough}.%
\footnote{Our previous estimate \cite{Tch096} of 2300 K looks out to be somewhat
exaggerated.%
} Using it in the high-temperature estimate for the critical temperature
eq. (\ref{eq:CriticalTemperaturesHighTemperature}) we obtain for
the highest critical temperature (that of the transition from the
Curie paramagnetic to the 1D-RVB state) the value $\theta_{c}^{\mathrm{crit}}=490$
K, which lies fairly above the decomposition temperature \cite{Dronskowski}.
In case the Q1D-RVB phase is assumed to be responsible for the Pauli
paramagnetism we take $J_{ac}=1400$ K as a plausible estimate. With
use of it the highest critical temeperature $\theta_{c,ac}^{\mathrm{crit}}=435$
K is as well high enough. In that sence these estimates are consistent.
For the exchange parameter $J_{a}$ we notice that its values is pretty
stable. Using eq. (\ref{eq:CriticalTemperaturesHighTemperature})
for the temperature $\theta_{c\rightarrow c,a}^{\mathrm{crit}}=95$
K of the pseudogap opening in the 1D-RVB phase, we arrive to a reasonable
value of $J_{a}=550$ K. On the other hand considering this critical
value as the critical temperature $\theta_{c,ac\rightarrow c,a,ac}^{\mathrm{crit}}$
given by eq. (\ref{eq:2OPto3OPCriticalTemperature}) yields $J_{a}=560$
K which shows pretty good consistency between two options of defining
the Pauli paramagnetic phase.

\subsection{Structure manifestations of transitions between RVB phases }

\subsubsection{Theory}

Previously \cite{Tch097,Tch099} we were able to relate the RVB OPs
and the lattice constants through \textquotedbl{}magnetostriction\textquotedbl{}
- the linear coupling of the structure parameters with the exchange
parameters. These moves absolutely apply in the present model with
three exchange parameters and respective OPs. We assume as previously
a linear relationship between the exchange parameters $J_{\tau}$
and geometry parameters $\rho_{\lambda}$: 
\begin{equation}
J_{\tau}=J_{\tau0}+\sum_{\lambda}J_{\tau,\lambda}^{\prime}\rho_{\lambda}.\label{eq:Magnetostriction}
\end{equation}
Following \cite{MisurkinOvchinnikov} we assume that zeroes of $\rho$'s
correspond to a hypothetical structure the CuNCN crystal would have
provided the exchange interactions $J_{\tau}$ are turned off. Deformation
of this hypothetical structure requires the elastic energy 
\begin{equation}
\frac{1}{2}\sum_{\mu\lambda}K_{\mu\lambda}\rho_{\mu}\rho_{\lambda}\label{eq:Elastic}
\end{equation}
for each nearest neighbor Cu-Cu pair. The observed geometry of the
crystal where the exchange interactions result in formation of one
of the RVB phases corresponds to the minimum with respect to $\rho_{\lambda}$
of the free energy eq. (\ref{eq:FreeEnergySimplified}) to which the
elastric energy eq. (\ref{eq:Elastic}) is added and the exchange
parameters are replaced according to eq. (\ref{eq:Magnetostriction})
: 
\begin{equation}
\frac{\partial F}{\partial\rho_{\lambda}}+\sum_{\mu\lambda}K_{\mu\lambda}\rho_{\mu}=0.\label{eq:MechanicEquilibrium}
\end{equation}
Using the special form of the RVB free energy $F$ (eq. (\ref{eq:FreeEnergySimplified}))
we arrive to the explicit expression 
\begin{equation}
\rho_{\mu}=\sum_{\lambda}\left(K\right)_{\mu\lambda}^{-1}\left(\sum_{\tau}A_{\tau}\zeta_{\tau}^{2}J_{\tau,\lambda}^{\prime}\right),\label{eq:StructureChange}
\end{equation}
- the sought relation between the RVB OPs and their structure manifestations
($A_{\tau}$ are numerical coefficients: 3 for $\tau=a,c$; 6 for
$\tau=ac$) which further generalizes the famous bond-length-bond-order
relation to the RVB states. Details are explained in Appendix \ref{sec:Theory-of-the}.

\subsubsection{Synchrotron measurements on CuNCN as explained by RVB phase transitions}

Now we can formulate what one could expect in the structural studies
provided CuNCN undergoes transitions between various RVB phases. The
vanishing OP's do not have any effect on the crystal structure. On
the other hand, a transition accompanied by splitting from zero of
some OP will be manifested in the structure changes as prescribed
by eq. (\ref{eq:StructureChange}). The sign of the effect is controlled
by that of the magnetostriction parameters $J_{\tau,\lambda}^{\prime}$.
The temperature dependence of the structure described by the parameters
$\rho_{\lambda}$ is thus that of the relevant combination of the
squares of the corresponding OPs, which can be different in different
phases. Previously \cite{Tch097,Tch099} we could relate the experimental
data on the anomalous temperature dependence of the lattice parameter
\emph{a} with the 1D-RVB to 2D-RVB phase transition of the anisoropic
triangular Heisenberg model as accompanied by the opening of the gap
in the quasiparticle spectrum in the \emph{a}-direction. However,
the experiment \cite{Tch099} showed some anomalies for the lattice
parameter \emph{c} at the temperature of the tentative 1D-RVB to 2D-RVB
phase transition ($80\div100$ K) and some more irregularities either
in the \emph{a}- or in the \emph{c}-directions at \emph{ca.} 30 K.
We also already mentioned above that the system of the exchange parameters
of the anisoropic triangular Heisenberg model was not particularly
intuitive. 

\begin{figure}
\includegraphics[bb=0bp 0bp 446bp 269bp,scale=0.9]{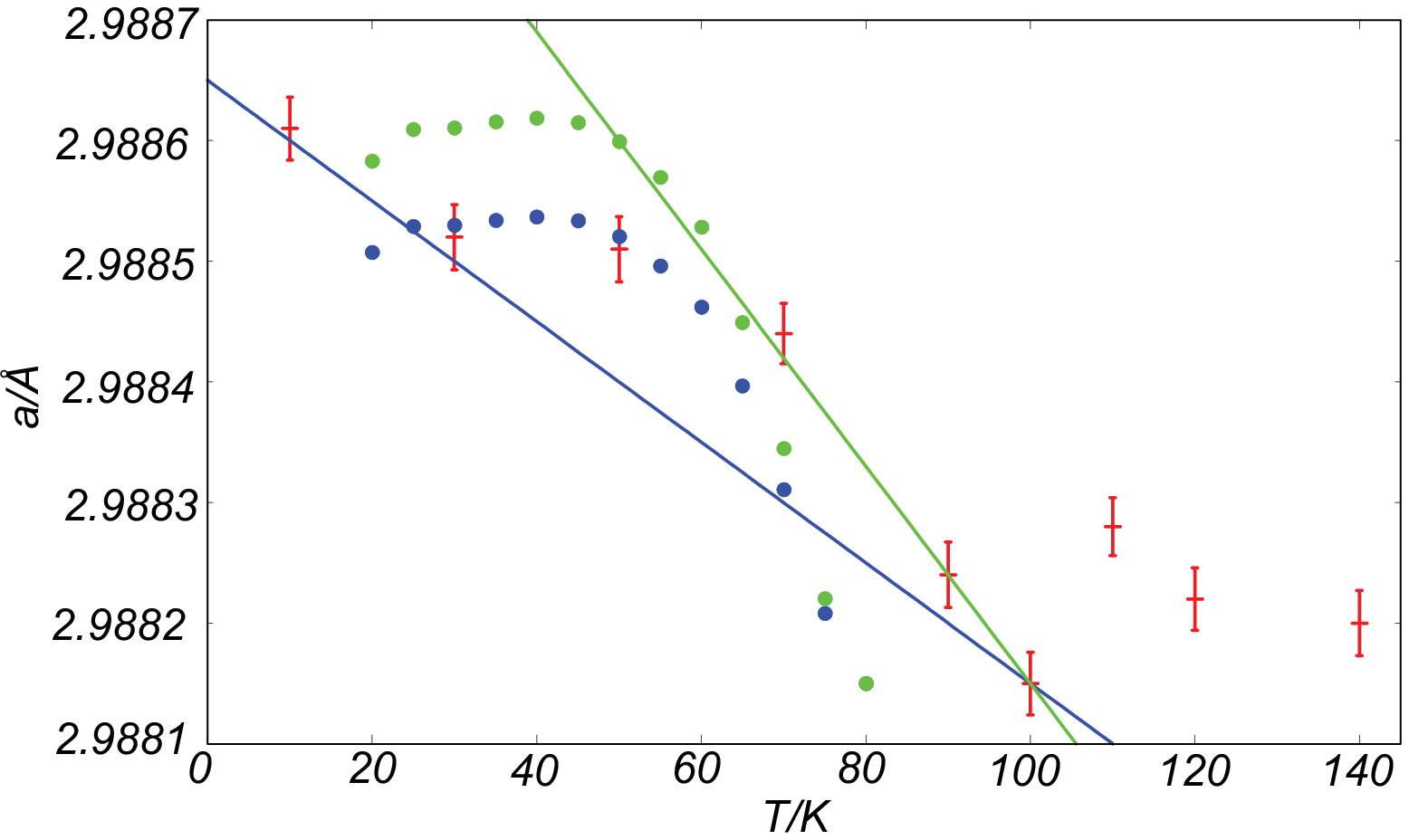}(a)

\includegraphics[scale=1.1]{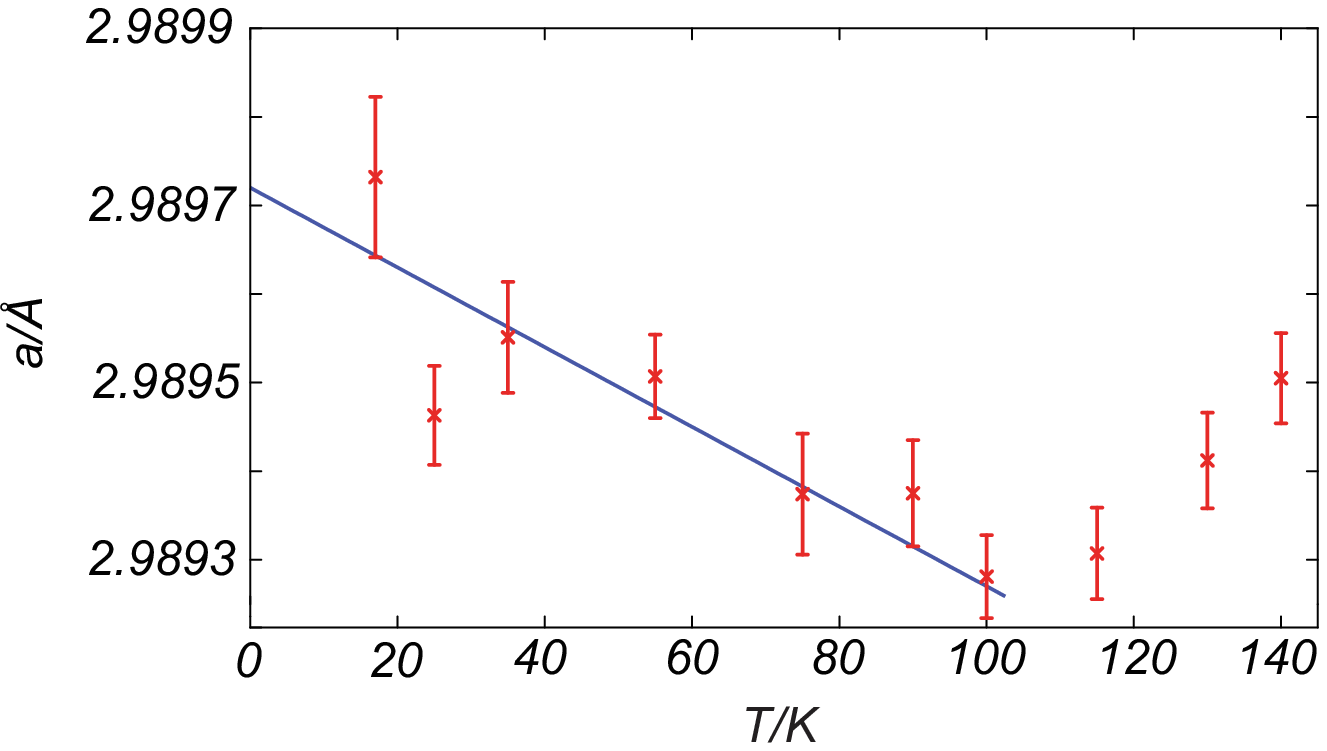}(b)

\caption{The temperature dependence of the $a$-lattice constant as extracted
(a) from synchrotron experiment\cite{Tch099} (red dashes with error
bars) as confronted with the $\zeta_{a}^{2}$ values extracted from
the ESR experiment and multiplied by a suitable constant $J_{a,a}^{\prime}/K_{aa}=0.08$
Å (blue dots) or confronted with the linear model eq. (\ref{eq:Lattice-a-parameter-vs-pseudogap})
with the parameters $\theta^{\mathrm{crit}}=100$ K, $\nu=0.5$, $A_{0}=140$
K extracted from one of the magnetic susceptibility fits; (b) from
the neutron experiment\cite{Houben-private} (red crosses with error
bars) and confronted with the linear model eq. (\ref{eq:Lattice-a-parameter-vs-pseudogap}).
\label{fig:Synchrotron-Neutron-a}}
\end{figure}

The results of the structural studeies related to the lattice parameter
\emph{a} are shown in Fig. \ref{fig:Synchrotron-Neutron-a}. 
We do not directly apply the general formula eq. (\ref{eq:StructureChange}),
but, first, take into account the approximately diagonal form of the
inverse matrix of the force constants as derived in Appendix \ref{sec:Estimate-of-the}
and thus conclude that the structure manifestations in the \emph{a}-
and \emph{c}-directions are independent. Since the exchange parameters
are sums of contributions of many superexchange paths one can expect
that they depend on all types of interatomic separations that is the
magnetostriction constants $J_{\tau,\lambda}^{\prime}$ with $\tau=a,c,ac$
and $\lambda=a,c$ are not vanishing for all combinations of $\tau$
and $\lambda$. It seems, however, to be reasonable, that the exchange
constant $J_{a}$ is not dependent on the deformation in the \emph{c}-direction
and thus $J_{a,c}^{\prime}=0$. Under these assumptions the structural
manifestations in the \emph{a}-direction decouple and can be recovered
by using $\tau,\lambda=a,\,\rho_{a}=\delta a=(a-a_{0})$ in eq. (\ref{eq:StructureChange}).
Like previously \cite{Tch097,Tch099} we conclude that the equilibrium
value of $\delta a$ in the RVB-phase with two pseudogaps must be
proportional to the squared OP $\zeta_{a}$. Due to its direct relation
with the pseudogap measured in the EPR experiements \cite{Tch096}
we can relate two observed quantities: the deformations and the temperature
dependent activation energy (the pseudogap \emph{$A(\theta)$}) in
the quasi-Arrhenius regime: 
\begin{equation}
\delta a=\frac{J_{a,a}^{\prime}}{3K_{a,a}J_{a}^{2}}A(\theta)^{2}\label{eq:Lattice-a-parameter-vs-pseudogap}
\end{equation}
The crucial point is the sign of the effect. At the first glance the
situation seems to be counterintuitive since in order to conform with
the experiment (increase of the $a$-parameter in the 2D-RVB phase
with two two-pseudogaps as compared to the (Q)1D-RVB phase (2, red
- or 5, magenta) the exchange parameter $J_{a}$ has to increase with
the increase of the Cu-Cu interatomic separation. However, the effective
value of the (antiferromagnetic) exchange parameter $J_{a}$ is a
sum of numerous contributions of different signs: 
\[
J_{a}=J_{a}(\mathrm{antiferro})-J_{a}(\mathrm{ferro})>0,
\]
where both $J_{a}(\mathrm{antiferro})$ and $J_{a}(\mathrm{ferro})$
are positive. The antiferromagnetic contribution is accumulated by
summing up contributions from numerous superexchange paths and is
weakly affected by the Cu-Cu interatomic separation since no direct
matrix elements between the states of the two Cu atoms affects $J_{a}(\mathrm{antiferro})$.
By contrast the ferromagnetic contribution strongly depends on the
angle $\mathrm{\widehat{CuNCu}}$ (see Refs. \cite{Goodenough,AtanasovAngelovMayer.})
and in the range of $\mathrm{\widehat{CuNCu}}>90^{\circ}$, which
is the case for CuNCN, decreases while the $\mathrm{\widehat{CuNCu}}$
angle and thus the Cu-Cu separation increases. Respectively the effective
antiferromagnetic exchange parameter $J_{a}$ increases while the
counterpoising ferromagnetic contribution decreases. This explains
the overall positive sign of $J_{a,a}^{\prime}$ and thus the experimental
fact of increase of the lattice parameter $a$ in the phase with the
pseudogap developing in the $a$-direction.

The quantitative agreement (shown in Fig. \ref{fig:Synchrotron-Neutron-a}(a)
by blue and green dots) is achieved by confronting the amplitude of
the structure effect $4\cdot10^{-4}$ Å \cite{Tch099} in the \emph{a}-direction
with the zero temperature limit of the pseudogap $A(\theta\rightarrow0)$
of \emph{ca.} 70 K \cite{Tch096}. The problem with the latter estimate
is that it is determined within somewhat different model \cite{Tch096,Tch097}.
Alternatively we take the value of $A_{0}=140$ K and the corresponding
classical value of the exponent $\nu=\frac{1}{2}$ as a plausible
estimate for the zero temperature limit for the pseudogap and with
the above value $J_{a}=560$ K we get $\zeta_{a}(\theta\rightarrow0)=0.08$.
Then the ratio of effective constants finally responsible for the
spin-phonon coupling is $J_{a,a}^{\prime}/K_{aa}=0.19$ Å. Combining
this with the value of the force constant $K_{aa}$ derived in Appendix
\ref{sec:Estimate-of-the} we can now evaluate $J_{a,a}^{\prime}$
to be \emph{ca.} 12000 K/Å which seems to fairly fit the expectations.

\begin{figure}
\includegraphics[bb=0bp 0bp 481bp 377bp,scale=0.8]{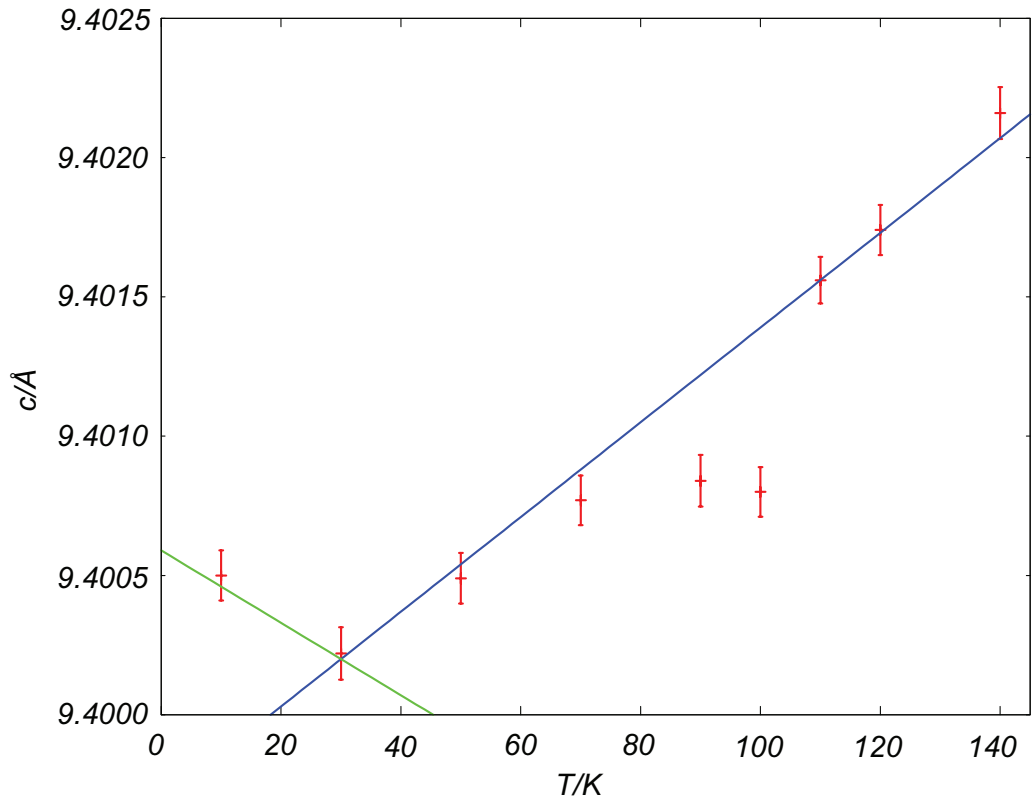}\caption{The temperature dependence of the $c$-lattice constant as extracted
from the synchrotron experiment \cite{Tch099}. \label{fig:The-temperature-dependence-synchrotron-c}}
\end{figure}

The synchrotron experiment was subject to a heavy and unfair criticism.
Thus the neutron difraction study of the temperature dependence of
the CuNCN structure have been performed \cite{Houben-private}. The
results concerning the \emph{a} lattice parameter are shown in Fig.
\ref{fig:Synchrotron-Neutron-a}(b). The most remarkable is that the
amplitude of the structure change as measured in the neutron scattering
coincides with that coming from the synchrotron one: $4\cdot10^{-4}$
Å. That means that our previous estimates of the parameters rationalizing
the synchrotron and ESR data remain valid. In Fig. \ref{fig:Synchrotron-Neutron-a}(a),
(b) the experimental data fairly lie on a straight line in the temperature
range between 100 K (the measured minimum of $a$ determined by the
neutron difraction and somewhat less characteristic for the synchrotron
data) and 30 K. This agrees with an assumption that in a rather wide
range the temperature dependence of the OPs follow the standard temperature
behavior with the classical value of the critical exponent $\nu=\frac{1}{2}$
since accoridng to our treatment eqs. (\ref{eq:StructureChange}),
(\ref{eq:Lattice-a-parameter-vs-pseudogap}) the lattice parameters
must obey the temperature evolution with the exponent $2\nu$ \emph{i.e.}
be linear, as observed. This, however, changes to somewhat more chaotic
behavior at \emph{ca.} 30 K. This incidentally corroborates with the
temperature run of the \emph{c} lattice parameter measured in the
synchrotron experiment \cite{Tch099} (Fig. \ref{fig:The-temperature-dependence-synchrotron-c})
which did not receive due attention so far. 
As one can see the lattice parameter \emph{c} generally (almost linearly)
decreases with temperature. It shows some irregularities (not changing
the sense of the run) about $80\div100$ K, where the magnetic susceptibilty
changes from the quasi-Pauli to the quasi-Arrhenius regime due to
opening the pseudogap in the \emph{a}-direction. However, the sense
of the temperature run of \emph{c} changes at 30 K where the lattice
parameter \emph{a} starts to show irregularities as well. 

In order to rationalize the low-temperature behavior of the lattice
parameter \emph{c} and eventually of \emph{a} we introduce one more
geometry variable $2\rho_{c}=\delta c=c-c_{0}$ and apply the general
relation eq. (\ref{eq:StructureChange}) and by this show that the
variation of two lattice parameters is: 
\begin{equation}
\left(\begin{array}{c}
\delta a\\
\delta c
\end{array}\right)=\frac{3}{\left|K\right|}\left(\begin{array}{c}
\left(K_{c,c}J_{a,a}^{\prime}-2K_{ac}J_{a,c}^{\prime}\right)\zeta_{a}^{2}+\left(K_{c,c}J_{c,a}^{\prime}-2K_{ac}J_{c,c}^{\prime}\right)\zeta_{c}^{2}\\
\left(-K_{ac}J_{a,a}^{\prime}+2K_{a,a}J_{a,c}^{\prime}\right)\zeta_{a}^{2}+\left(-K_{ac}J_{c,a}^{\prime}+2K_{a,a}J_{c,c}^{\prime}\right)\zeta_{c}^{2}
\end{array}\right),\label{eq:C-Lattice-vs-OPs}
\end{equation}
where we denote by $\left|K\right|$ the determinant of the 2x2 matrix
of the force constants and make use of our previous consideration
allowing to neglect the OP $\zeta_{ac}$. 

In order to qualitatively understand the temperature run of the lattice
parameters \emph{a} and \emph{c} formally given by eq. (\ref{eq:C-Lattice-vs-OPs})
we take a closer look at the lower-temperature range of the parameter
phase diagram Fig. \ref{Flo:ParameterPhaseDiagrams}. There one sees
that below the octal point the thermal evolution consists \emph{en
gros }in squeezing out all the phases by the 2D-RVB phase with vanishing
$\zeta_{ac}$ (7 - orange). The 2D-RVB phase with three nonvanishing
OP's (8 - yellow) is transient, its parameters' area is never large
and it is subject to deformations and displacements under the {}``pressure''
of the phase 7. The same {}``pressure'' squeezes the area of existence
of the Q1D-RVB phase 5 (magenta) as well. Assuming the position of
CuNCN on the parameters phase diagram Fig. \ref{Flo:ParameterPhaseDiagrams}
within the magenta phase, but close and somewhat below the quadruple
point of the phases 2, 5, 7, and 8 (red, magenta, orange, and yellow)
in the rightmost graph in the middle row one can see on the following
slices the following sequence of transitions between the RVB phases:
\[
\begin{array}{ccccc}
\mathrm{Q1D-RVB} & \rightarrow & \mathrm{2D-RVB} & \rightarrow & \mathrm{2D-RVB}\\
\zeta_{c},\zeta_{ac} & \rightarrow & \zeta_{c},\zeta_{ac},\zeta_{a} & \rightarrow & \zeta_{c},\zeta_{a}\\
\mathrm{magenta} & \rightarrow & \mathrm{yellow} & \rightarrow & \mathrm{orange}
\end{array}.
\]
The thermal dependence of the OPs \emph{within} these phases is described
by the formulae of Table \ref{tab:OPsTemeperatureDependenceCuNCN}.
One should not expect that these 
\begin{table}
\caption{Temeperature dependencies of the OPs for possible phases of CuNCN
in the high-temperature approximation of the \emph{c}-\emph{a}-\emph{ca}-RVB
model. The areas of the existence of the corresponding phases are
given by the condition of positiveness of the expressions under the
square roots. \label{tab:OPsTemeperatureDependenceCuNCN} }
\begin{tabular}{|l|l|}
\hline 
Color code & OP's \emph{vs. $\theta$}\tabularnewline
\hline 
\hline 
Magneta & $\begin{cases}
\zeta_{a}=0\\
\zeta_{c}=\frac{4\theta}{3J_{c}}\sqrt{1-\theta/\theta_{c,ac}^{\mathrm{crit}}} & \theta_{c,ac}^{\mathrm{crit}}=\frac{3J_{ac}J_{c}}{8\left(3J_{ac}-2J_{c}\right)}\\
\zeta_{ac}=\frac{8}{3}\sqrt{\frac{2}{3}}\frac{\theta}{J_{ac}}\sqrt{\frac{\theta\left(J_{ac}-J_{c}\right)}{J_{ac}J_{c}}}
\end{cases}$\tabularnewline
\hline 
Yellow & $\begin{cases}
\zeta_{a}=\frac{4\theta}{3J_{a}}\sqrt{1-\theta/\theta_{c,ac\rightarrow a,c,ac}^{\mathrm{crit}}} & \theta_{c,ac\rightarrow a,c,ac}^{\mathrm{crit}}=\frac{3J_{ac}J_{a}}{8\left(3J_{ac}-2J_{a}\right)}\\
\zeta_{c}=\frac{4\theta}{3J_{c}}\sqrt{1-\theta/\theta_{a,ac\rightarrow a,c,ac}^{\mathrm{crit}}} & \theta_{a,ac\rightarrow a,c,ac}^{\mathrm{crit}}=\frac{3J_{ac}J_{c}}{8\left(3J_{ac}-2J_{c}\right)}\\
\zeta_{ac}=\frac{4}{3}\sqrt{\frac{1}{3}}\frac{\theta}{J_{ac}}\sqrt{\theta/\theta_{a,c,ac}^{\mathrm{crit}}-1} & \theta_{a,c,ac}^{\mathrm{crit}}=\mathrm{eq.\,}(\ref{eq:3OPsCriticalTemperature})
\end{cases}$\tabularnewline
\hline 
Orange & $\begin{cases}
\zeta_{a}=\frac{4\theta}{\sqrt{15}J_{a}}\sqrt{1-\theta/\theta_{c\rightarrow a,c}^{\mathrm{crit}}} & \theta_{c\rightarrow a,c}^{\mathrm{crit}}=\frac{3J_{a}J_{c}}{8\left(3J_{c}-2J_{a}\right)}\\
\zeta_{c}=\frac{4\theta}{\sqrt{15}J_{c}}\sqrt{1-\theta/\theta_{a\rightarrow a,c}^{\mathrm{crit}}} & \theta_{a\rightarrow a,c}^{\mathrm{crit}}=\frac{3J_{a}J_{c}}{8\left(3J_{a}-2J_{c}\right)}\\
\zeta_{ac}=0
\end{cases}$\tabularnewline
\hline 
\end{tabular}
\end{table}
expressions derived from the high-temperature expansion for the free
energy are exactly valid at low-temperature. Specifically, the prefactors
$\theta$ should not be taken seriously since we expect that the $\zeta_{a,c}$
OPs flow to some finite values as temperature flows to zero, although
splitting from zero vaues at the critical temperatures given next
to them. Upto that uncertain factor the temperature dependence of
the OPs in three presumably observed phases is as follows. As we explained
earlier, since the exchange parameters satisfy the condition $J_{ac}\gtrapprox J_{c}$
the OP $\zeta_{ac}$ is always small and decreases with decreasing
temperature, thus we do not consider it explicitly further. The OP
$\zeta_{c}$ should in principle increase either in the magneta or
in yellow phases, but we assume that in the interesting temperature
range $\theta\ll\theta_{c,ac}^{\mathrm{crit}}=\theta_{a,ac\rightarrow a,c,ac}^{\mathrm{crit}}=435$
K it has almost reached its zero temperature limit and does not significntly
change any more.%
\footnote{Possible minor effect of increase of $\zeta_{c}$ produces no visible
temperature dependence of the Q1D-RVB quasiparticle bandwidth due
to decrease of $\zeta_{ac}$ since these two contributions changing
in opposite directions compensate variations of each other. %
} The OP $\zeta_{a}$ splits from zero at the critical temperature
of $\theta_{c,ac\rightarrow a,c,ac}^{\mathrm{crit}}\approx100$ K
(transition to the yellow phase) which affects the \emph{a} lattice
parameter as explained above. In the general setting eq. (\ref{eq:C-Lattice-vs-OPs}),
but under previous assumption of $J_{a,c}^{\prime}=0$ the temeperature
independent $\zeta_{c}^{2}$ does not contribute to the visible variation
of the lattice constants. However, the lattice parameter \emph{c}
turns out to be sensitive to the pseudogap opening in the \emph{a}
direction through the off-diagonal element of the the inverse matrix
of the force constants. Particularly remarkable is the fact that the
effect on the \emph{c} is predicted to have the sign opposite to that
on \emph{a} (\emph{a} increases, \emph{c} decreases), as observed.
Finally at the lowest accessible critical temperature $\theta_{a,c,ac}^{\mathrm{crit}}$
eq. ($\ref{eq:3OPsCriticalTemperature}$) a transition to the orange
phase takes place. A word of caution needs to be said here: the present
set of the exchange parameters yields very narrow temeperature range
where the transient (yellow) phase can exist. We assume that eq. ($\ref{eq:3OPsCriticalTemperature}$)
strongly overestimates this temperature (92 K), which must by considerably
lower (\emph{ca.} 30 K) since one cannot rely upon the results of
the high-temperature expansion any more. For obtaining more reliable
estimates one needs to know at least the zero temperature limit of
the \emph{a}-\emph{c}-\emph{ca} model which is yet to be done. However,
we assume that the general shape of the phase diagram is correctly
reproduced by the high-temperature expansion and proceed within this
setting. The evanscence of $\zeta_{ac}$ at $\theta_{a,c,ac}^{\mathrm{crit}}$
whatever it is affects neither the bandwidth, since this OP is never
large, nor the character of the temperature dependence of $\zeta_{a}$,
although the variation of the slope can be as well expected. However,
at this phase transition a remarkable change can be expected to the
character of the temperature dependence of $\zeta_{c}$. Namely, it
switches from increase to decrease. This happens because of an instanteneous
change of its reference temperature shown in Table \ref{tab:OPsTemeperatureDependenceCuNCN}
from positive $\theta_{a,ac\rightarrow a,c,ac}^{\mathrm{crit}}$ to
$\theta_{a\rightarrow a,c}^{\mathrm{crit}}$ which is \emph{negative}
(and much smaller by absolute value: \emph{ca.} $-300$ K) in the
relevant area of the exchange parameters' ($J_{ac}\gtrapprox J_{c}>J_{a}$)
space. Together changing the the sign and the magnitide ot the reference
temperature causes the change of the temperature run of the OP $\zeta_{c}$
which starts to decrease in the orange phase and through eq. (\ref{eq:C-Lattice-vs-OPs})
affect the temperature behavior of the lattice parameters \emph{a}
and $c$. This tentatively and qualitatively explains the anomalies
observed in Figs. \ref{fig:Synchrotron-Neutron-a} and \ref{fig:The-temperature-dependence-synchrotron-c}
although the available amount of experimental data does not suffice
to reliably estimate quite a few magnetostriction and other parameters
required for at least semiquantitative description. In the minimal
setting: \emph{i.e.} neglecting the off-diagonal elements of the inverse
matrix of force constants and the off-diagonal magnetostriction terms
we immeditely obtain 
\[
\delta c=\frac{6J_{c,c}^{\prime}}{K_{c,c}}\zeta_{c}^{2}
\]
which allows to at least conclude that the magnetostriction parameter
$J_{c,c}^{\prime}$ has intuitively understandable sign: it is negative
since $\delta c$ turns to be positive when the variation of OP $\zeta_{c}$
is negative.

\section{Conclusion}

A new form of the frustrated spatially anisotropic antiferromagnetic
Heisenberg Hamiltonian close to the popular $J_{1}J_{2}J_{3}$ model
with exchange parameters $J_{c},\, J_{a},\,\mathrm{and}\, J_{ac}$
extended along the $c$, $a$, and $a\pm c$ directions of a two-dimensional
rectangular lattice is proposed. When applied to model fascinating
physics of copper carbodiimide (CuNCN) it explains the absence of
magnetic order in CuNCN down to 4 K by assuming resonanting valence
bond (RVB) character of the emerging phases. The quasiparticle spectrum
of the RVB model of the proposed Hamiltonian shows three principal
regimes: (i) a state with two pairs of lines of nodes, (ii) states
with a pair of lines of nodes (termed as 1D- and Q1D-RVB states),
(iii) states with two pseudogaps and four nodal points (2D-RVB states).
Extraordinary rich parameters-temperature phase diagram of the model
contains eight different phases whose magnetic behavior includes Curie
and quasi-Pauli paramagnetism (1D- and Q1D-RVB phases), and (pseudo)gapped
(quasi-Arrhenius) paramagnetism (2D-RVB phases). Adding magnetostriction
and elastic terms to the free energy of the model explains the the
temperature dependence of the CuNCN crystal structure by assuming
that a sequence of transitions between different RVB phases occurs
in CuNCN while temperature decreases. Confronting the model with the
magnetic susceptibility and strucutre (both synchrotron and neutron)
data recorded as functions of temperature in the range between $ca.$
10 and 200 K shows a remarkably good agreement between the theoretical
predictions and the experiment which is reached by ascribing the values
of the model parameters which are intuitively acceptable both in terms
of their absolute magnitudes, relative values, and the character of
their geometry dependence.

\section*{Acknowledgments}

This work has been performed with the support of Deutsche Forschungsgemeinschaft.
In addition, we acknowledge the Russian Foundation for Basic Research
for the financial support dispatched to ALT through the grant No.
13-03-90430. Dr. Andrej Zorko of Jožef Stefan Institute (Ljubljana,
Slovenia) is acknowledged for sending numerical data on the $T$-dependence
of the energy gap as extracted from the ESR measurements \cite{Tch096}
as well as of the data on the ESR susceptibility. Prof. Dr. U. Ruschewitz
of the University of Cologne is acknowledged for sending numerical
data on the $T$-dependence of the lattice parameters as drived from
the synchrotron experiments \cite{Tch099}.

\newpage{}

\appendix

\section{Equations of motion and self consistency equations.\label{sec:Equations-of-motion} }

Equations of motion are based on the Heisenberg representation in
which each operator obeys the following: 
\begin{equation}
i\hbar\dot{A}=\left[A,H\right]\label{eq:HeisenbergEOM-1}
\end{equation}
where $\left[...,...\right]$ stands for the commutator and \textquotedbl{}$\dot{}$\textquotedbl{}
for the time derivative. Applying this to the creation and annihilation
operators $c_{\mathbf{r}\sigma}^{+}(c_{\mathbf{r}\sigma})$ and performing
commutation, mean field decoupling and Fourier transformation as done
previously \cite{Tch091} results in mean field equations of motion
for them: 
\begin{eqnarray}
i\hbar\dot{c}_{\mathbf{k}\sigma}=-\frac{3}{2}\sum_{\mathbf{\tau}}J_{\mathbf{\boldsymbol{\tau}}}\xi_{\mathbf{\boldsymbol{\tau}}}\cos(\mathbf{k\tau})c_{\mathbf{k}\sigma} & -\frac{3}{2}\sum_{\mathbf{\tau}}J_{\mathbf{\boldsymbol{\tau}}}\Delta_{\boldsymbol{\tau}} & \cos(\mathbf{k\tau})c_{-\mathbf{k}-\sigma}^{+}\nonumber \\
i\hbar\dot{c}_{\mathbf{k}\sigma}^{+}=\frac{3}{2}\sum_{\mathbf{\tau}}J_{\mathbf{\boldsymbol{\tau}}}\xi_{\mathbf{\boldsymbol{\tau}}}\cos(\mathbf{k\tau})c_{\mathbf{k}\beta}^{+} & -\frac{3}{2}\sum_{\mathbf{\tau}}J_{\mathbf{\boldsymbol{\tau}}}\Delta_{\boldsymbol{\tau}}^{*} & \cos(\mathbf{k\tau})c_{-\mathbf{k}-\sigma}\label{eq:MeanFieldEOM-1}
\end{eqnarray}
 These latter reduces to the set of $2\times2$ eigenvalue problems
for each wave vector $\boldsymbol{\mathbf{k}}$:

\[
\left(\begin{array}{cc}
\xi_{\mathbf{k}} & \Delta_{\mathbf{k}}\\
\Delta_{\mathbf{k}}^{*} & -\xi_{\mathbf{k}}
\end{array}\right)\left(\begin{array}{c}
u_{\mathbf{k}}\\
v_{\mathbf{k}}
\end{array}\right)=E_{\mathbf{k}}\left(\begin{array}{c}
u_{\mathbf{k}}\\
v_{\mathbf{k}}
\end{array}\right)
\]
 with 
\begin{eqnarray*}
\xi_{\mathbf{k}} & = & -3\sum_{\mathbf{\tau}}J_{\mathbf{\boldsymbol{\tau}}}\xi_{\mathbf{\boldsymbol{\tau}}}\cos(\mathbf{k\mathbf{\tau}})\\
\Delta_{\mathbf{k}} & = & 3\sum_{\mathbf{\tau}}J_{\mathbf{\boldsymbol{\tau}}}\Delta_{\mathbf{\boldsymbol{\tau}}}\cos(\mathbf{k\mathbf{\tau}})
\end{eqnarray*}
 (summation over $\mathbf{\boldsymbol{\tau}}$ extends to $\mathbf{\tau}_{i};i=1\div4$)
which results in the eigenvalues:

\begin{eqnarray*}
E_{\mathbf{k}} & =\pm & \sqrt{\xi_{\mathbf{k}}^{2}+\left|\Delta_{\mathbf{k}}\right|^{2}}
\end{eqnarray*}
 whose eigenvectors are combinations of the annihilation and creation
operators with the coefficients $u_{\mathbf{k}},v_{\mathbf{k}}$.
This set of equations closes by the selfconsistency conditions: 
\begin{eqnarray}
\xi_{\mathbf{\boldsymbol{\tau}}} & = & -\frac{1}{2N}\sum_{\mathbf{k}}\exp(i\mathbf{k\tau})\frac{\xi_{\mathbf{k}}}{E_{\mathbf{k}}}\tanh\left(\frac{E_{\mathbf{k}}}{2\theta}\right)\nonumber \\
\Delta_{\mathbf{\boldsymbol{\tau}}} & = & \frac{1}{2N}\sum_{\mathbf{k}}\exp(-i\mathbf{k\tau})\frac{\Delta_{\mathbf{k}}}{E_{\mathbf{k}}}\tanh\left(\frac{E_{\mathbf{k}}}{2\theta}\right)\label{eq:SelfConsistencyEquations-1}
\end{eqnarray}
 for the order parameters (OPs). The lattice symmetry considerations
allow us to restrict ourselves by the OPs: $\xi_{a},\xi_{c},\xi_{ac};\Delta_{a},\Delta_{c},\Delta_{ac}$.
For the complex OPs we introduce a polar representation:

\noindent 
\[
\Delta_{\mathbf{\boldsymbol{\tau}}}=\eta_{\mathbf{\boldsymbol{\tau}}}e^{i\varphi_{\mathbf{\tau}}}.
\]
 The standard moves foreseen for the $SU(2)$ symmetric solutions
are used to exclude the cross terms in OPs from $E_{\mathbf{k}}^{2}$
which leads to the system of equations: 
\[
\xi_{\mathbf{\boldsymbol{\tau}}}\xi_{\mathbf{\boldsymbol{\tau}^{\prime}}}=-\eta_{\mathbf{\boldsymbol{\tau}}}\eta_{\mathbf{\boldsymbol{\tau}^{\prime}}}\cos(\varphi_{\mathbf{\tau}}-\varphi_{\mathbf{\tau}^{\prime}})
\]
 (three equations). Introducing the relative phases as: $\theta_{a}=\varphi_{a}-\varphi_{ac}$;
$\theta_{c}=\varphi_{c}-\varphi_{ac}$ we arrive to the equations
of the form:%
\footnote{\noindent The consequences of setting $\theta_{c}=\varphi_{ac}-\varphi_{c}$
and thus having the first equation in the form $\xi_{a}\xi_{c}+\eta_{a}\eta_{c}\cos(\theta_{a}+\theta_{c})=0$,
which exactly coincides with \cite{Ogata2003}, yet have to be studied.
Most probably it brings up more degenerate phases with various phase
angles.%
} 
\begin{eqnarray*}
\xi_{a}\xi_{c}+\eta_{a}\eta_{c}\cos(\theta_{a}-\theta_{c}) & = & 0\\
\xi_{a}\xi_{ac}+\eta_{a}\eta_{ac}\cos\theta_{a} & = & 0\\
\xi_{c}\xi_{ac}+\eta_{c}\eta_{ac}\cos\theta_{c} & = & 0
\end{eqnarray*}
 similar to those derived in \cite{Ogata2003} which can be satisfied
\emph{e.g.} by setting 
\begin{eqnarray}
\xi_{a}\xi_{c} & =- & \eta_{a}\eta_{c}\neq0\nonumber \\
\theta_{a} & =\theta_{c} & =\frac{\pi}{2}\label{eq:PossibleRVBSolutions-1}\\
\eta_{ac} & \neq & 0;\xi_{ac}=0\nonumber 
\end{eqnarray}
 Of course, the OPs can be also vanishing.

Under the above conditions the spectrum of quasiparticles acquires
the form: 
\begin{eqnarray}
E_{\mathbf{k}}^{2} & = & 9\left(J_{a}^{2}\left(\xi_{a}^{2}+\eta_{a}^{2}\right)\cos^{2}x+J_{c}^{2}\left(\xi_{c}^{2}+\eta_{c}^{2}\right)\cos^{2}z+4J_{ac}^{2}\left(\xi_{ac}^{2}+\eta_{ac}^{2}\right)\cos^{2}x\cos^{2}z\right),\label{eq:Spectrum-1}
\end{eqnarray}
 where we set $x=\mathbf{k}_{x};z=\mathbf{k}_{z}.$

\section{High-temperature expansion.\label{sec:High-temperature-expansion.}}

At high temperature we can use an expansion: 
\[
\ln\left(2\cosh\left(\frac{E_{\mathbf{k}}}{2\theta}\right)\right)\approx\ln2+\frac{1}{2}\left(\frac{E_{\mathbf{k}}}{2\theta}\right)^{2}-\frac{1}{12}\left(\frac{E_{\mathbf{k}}}{2\theta}\right)^{4}
\]
 which when inserted in eq. (\ref{eq:FreeEnergySimplified}) integrates
explicitly. For determining the critical temperatures to the first
approximation it suffies to restrict ourselves by the second power
terms. This results in an expression quadratic in $\zeta_{\tau}$.
Combining thus obtained \textquotedbl{}kinetic\textquotedbl{} energy:

\[
-2\theta\cdot\frac{9}{8\theta^{2}}\left(\frac{1}{2}J_{a}^{2}\zeta_{a}^{2}+\frac{1}{2}J_{c}^{2}\zeta_{c}^{2}+J_{ac}^{2}\zeta_{ac}^{2}\right),
\]
 with the potential energy terms from eq. (\ref{eq:FreeEnergySimplified})
we get: 
\[
F_{\mathrm{HT}}=\sum_{\tau}\left[-\frac{9}{8\theta}J_{\tau}^{2}+3J_{\tau}\right]\zeta_{\tau}^{2}.
\]
This result can be improved with use of the Ginzburg-Landau approximate
free energy $F_{\mathrm{GL}}(\zeta_{\tau},\theta)$ which involves
the terms up to the fourth power in $\zeta_{\tau}$'s. They appear
from the integration of the 4-th power of the spectrum which performs
explicitly and yields the \textquotedbl{}kinetic\textquotedbl{} energy
of the form:
\begin{align*}
\frac{1}{768\theta^{3}} & \left[36\zeta_{a}^{2}J_{a}^{2}\left(27\zeta_{ac}^{2}J_{ac}^{2}+9\zeta_{c}^{2}J_{c}^{2}-24\theta^{2}\right)\right.\\
+ & 243\zeta_{a}^{4}J_{a}^{4}+3\left(36\zeta_{ac}^{2}J_{ac}^{2}\left(9\zeta_{c}^{2}J_{c}^{2}-16\theta^{2}\right)\right)\\
+ & \left.486\zeta_{ac}^{4}J_{ac}^{4}-288\theta^{2}\zeta_{c}^{2}J_{c}^{2}+81\zeta_{c}^{4}J_{c}^{4}\right]
\end{align*}
which together with the potential energy yields the free energy $F_{\mathrm{GL}}(\zeta_{\tau},\theta)$
used for further analysis.

\section{Quasiparticle densities of states in various RVB phases\label{sec:Quasiparticle-densities-of}}

In Section \ref{sub:Free-energy-and} we gave an impression of the
complexity of the phase diagram of the RVB model with three exchange
parameters. We also gave a brief description of the most characteristic
features of the qDoS in various RVB phases. Here we provide a brief
sketch of the derivation of qDoS given in Table \ref{tab:OPsTemeperatureDependence}. 

The definition of the qDoS reads: 
\[
g(\varepsilon)=\frac{1}{4\pi^{2}}\intop_{BZ}\delta\left(\varepsilon-E_{\mathbf{k}}\right)d^{2}\mathbf{k}
\]
Following Ref. \cite{Nagaev-density} we insert the integral representation
for the Dirac $\delta$-function: 
\begin{align*}
g(\varepsilon) & =\frac{1}{2\pi}\frac{1}{4\pi^{2}}\intop_{BZ}\intop_{-\infty}^{\infty}dt\exp\left(it\left(\varepsilon-E_{\mathbf{k}}\right)\right)d^{2}\mathbf{k}=\\
= & \frac{1}{2\pi}\intop_{-\infty}^{\infty}dt\left[\exp\left(it\varepsilon\right)\frac{1}{4\pi^{2}}\intop_{BZ}\exp\left(-itE_{\mathbf{k}}\right)d^{2}\mathbf{k}\right].
\end{align*}
 For all phases having lines of nodes (those with numbers 2 - 6) the
integration over one of the components of the wave vector $\mathbf{k}$
in the BZ is performed and yields an intermediate result in terms
of the the Bessel and Struve functions of arguments dependent on the
Fourier transformation variable $t$ and the remaining component of
the wave vector $\mathbf{k}$. The Fourier trasnforms with respect
to $t$ can be done for the intermediate answers of that form. It
yields integrands of the elliptic integrals over the remaning component
of the wave vector $\mathbf{k}$. This solves the problem of calculationg
the qDoS for the dispersion laws with the lines of nodes in the BZ.
The results are given in respective cells of Table \ref{tab:OPsTemeperatureDependence}.

In order to cope with remaining two phases whose spectra contain only
nodal points (numbers 7 and 8) one more trick, namely performing previous
moves for the squared spectrum and thus obtaining the distribution
of the states as a function of their squared energy solves the problem.
The distribution of squares of the quasiparticle energies is: 
\[
\varrho(\varepsilon^{2})=\frac{1}{4\pi^{2}}\intop_{BZ}\delta\left(\varepsilon^{2}-E_{\mathbf{k}}^{2}\right)d^{2}\mathbf{k}.
\]
Then the sought qDoS is given by \cite{Tutubalin-book}: 
\[
g\left(\varepsilon\right)=2\varepsilon\varrho(\varepsilon^{2}).
\]
To obtain $\varrho(\varepsilon^{2})$ we again use the integral representation
of the Dirac $\delta$-function: 
\begin{eqnarray*}
\varrho(\varepsilon^{2}) & = & \frac{1}{2\pi}\frac{1}{4\pi^{2}}\intop_{BZ}\intop_{-\infty}^{\infty}dt\exp\left(it\left(\varepsilon^{2}-E_{\mathbf{k}}^{2}\right)\right)d^{2}\mathbf{k}=\\
 & = & \frac{1}{2\pi}\intop_{-\infty}^{\infty}dt\left[\exp\left(it\varepsilon^{2}\right)\frac{1}{4\pi^{2}}\intop_{BZ}\exp\left(-itE_{\mathbf{k}}^{2}\right)d^{2}\mathbf{k}\right].
\end{eqnarray*}
 Remarkably enough sequential integrations of the squared spectrum
over one of the components of the wave vector $\mathbf{k}$ and Fourier
transformation with respect to $t$ yield the expressions of the same
form as integration of the spectrum itself: the Bessel function and
an integrand of the elliptic integral. Thus the final intergation
over the remainig component of the wave vector $\mathbf{k}$ yields
some elliptic integrals given in respective cells of Table \ref{tab:OPsTemeperatureDependence}.

\section{Theory of the structural manifestations of the RVB states.\label{sec:Theory-of-the}}

We start from the mechanic equilibrium condition of the crystal in
a RVB phase: 
\begin{equation}
\frac{\partial F}{\partial\rho_{\lambda}}+\sum_{\mu\lambda}K_{\mu\lambda}\rho_{\mu}=0.\label{eq:MechanicEquilibrium-1}
\end{equation}
and notice that the free energy $F$ given by eq. (\ref{eq:FreeEnergySimplified})
has a special form. The first term ({}``kinetic'' energy) is integral
of a function of the dispersion law $E_{\mathbf{k}}$ whose argument
has the form: 
\[
\sum_{\tau}A_{\tau}^{2}J_{\tau}^{2}\zeta_{\tau}^{2}f_{\tau}^{2}\left(\mathbf{k}\right),
\]
 where $A_{\tau}$ are numerical coefficients (3 for $\tau=a,c$;
6 for $\tau=ac$); $f_{\tau}\left(\mathbf{k}\right)$ are trigonometrical
expressions ($\cos\mathbf{k}_{\tau}$ for $\tau=a,c$; $\cos\mathbf{k}_{a}\cos\mathbf{k}_{c}$
for $\tau=ac$). The {}``potential'' energy contribution to the
free energy eq. (\ref{eq:FreeEnergySimplified}) is: 
\[
\sum_{\tau}A_{\tau}J_{\tau}\zeta_{\tau}^{2}.
\]
Due to the above special form of the {}``kinetic'' and {}``potential''
energies the self-consistency equations for $\zeta_{\tau}$ have the
form: 
\begin{align*}
\intop_{BZ}\tanh\left(\frac{E_{\mathbf{k}}}{2\theta}\right)E_{\mathbf{k}}^{\prime}A_{\tau}^{2}J_{\tau}^{2}\zeta_{\tau}f_{\tau}^{2}\left(\mathbf{k}\right)d^{2}\mathbf{k} & =A_{\tau}J_{\tau}\zeta_{\tau}
\end{align*}
each, to be solved simultaneously for all $\tau$. Apparently whatever
(sub)set of $\zeta_{\tau}=0$ satisfies the equations. For the nonvanishing
OPs the self-consistency equations acquire the form:

\begin{equation}
\intop_{BZ}\tanh\left(\frac{E_{\mathbf{k}}}{2\theta}\right)E_{\mathbf{k}}^{\prime}f_{\tau}^{2}\left(\mathbf{k}\right)d^{2}\mathbf{k}=\frac{1}{A_{\tau}J_{\tau}}.\label{eq:SelfConsistency}
\end{equation}

Now we can turn to the structure determination. The derivative of
the kinetic energy with respect to the geometry parameters $\rho_{\lambda}$
reads:

\[
-2\sum_{\tau}A_{\tau}^{2}\left(J_{\tau}\zeta_{\tau}^{2}J_{\tau,\lambda}^{\prime}+J_{\tau}^{2}\zeta_{\tau}\frac{\partial\zeta_{\tau}}{\partial\rho_{\lambda}}\right)\intop_{BZ}\tanh\left(\frac{E_{\mathbf{k}}}{2\theta}\right)E_{\mathbf{k}}^{\prime}f_{\tau}^{2}\left(\mathbf{k}\right)d^{2}\mathbf{k}.
\]
The value of the above integral for the equilibrium values of the
OPs is given by the self-consistency conditions eq. (\ref{eq:SelfConsistency}).
Thus the derivative of the kinetic energy rewrites:
\[
-2\sum_{\tau}A_{\tau}\left(\zeta_{\tau}^{2}J_{\tau,\lambda}^{\prime}+J_{\tau}\zeta_{\tau}\frac{\partial\zeta_{\tau}}{\partial\rho_{\lambda}}\right).
\]
Combining this with the derivative of the potential enenrgy and that
of the elastic energy we get:
\[
-2\sum_{\tau}A_{\tau}\left(\zeta_{\tau}^{2}J_{\tau,\lambda}^{\prime}+J_{\tau}\zeta_{\tau}\frac{\partial\zeta_{\tau}}{\partial\rho_{\lambda}}\right)+\sum_{\tau}A_{\tau}\left(\zeta_{\tau}^{2}J_{\tau,\lambda}^{\prime}+2J_{\tau}\zeta_{\tau}\frac{\partial\zeta_{\tau}}{\partial\rho_{\lambda}}\right)+\sum_{\mu\lambda}K_{\mu\lambda}\rho_{\mu}=0.
\]
The terms including the derivative $\frac{\partial\zeta_{\tau}}{\partial\rho_{\lambda}}$
stemming from the kinetic and potential energies cancel each other:
a remarkable consequence of the Hellmann-Feynman and virial theorems,
which immediately results in:

\[
-\sum_{\tau}A_{\tau}\left(\zeta_{\tau}^{2}J_{\tau,\lambda}^{\prime}\right)+\sum_{\mu\lambda}K_{\mu\lambda}\rho_{\mu}=0,
\]
which after some trivial algebra results in:

\begin{equation}
\rho_{\mu}=\sum_{\lambda}\left(K\right)_{\mu\lambda}^{-1}\left(\sum_{\tau}A_{\tau}\zeta_{\tau}^{2}J_{\tau,\lambda}^{\prime}\right).\label{eq:StructureChange-1}
\end{equation}

\section{Estimate of the force matrix from the elastic constants\label{sec:Estimate-of-the}}

Now we notice that the deformation tensor (for methods used for this
and further evaluates see Ref. \cite{LL7}) corresponding to the structure
variation as desribed by the geometry deformation parameter $\rho_{a}=\delta a$
has only one nonvanishing component $u_{aa}=\delta a/a_{0}$. The
elastic energy of the unit cell under such deformation is: 
\[
\frac{1}{2}C_{aa,aa}a_{0}b_{0}c_{0}\left(\frac{\delta a}{a_{0}}\right)^{2}=4\times\frac{1}{2}K_{aa}\left(\delta a\right)^{2},
\]
 where $C_{aa,aa}$ is the corresponding element of the elasticity
moduli tensor, the multiplier of 4 in the right hand side takes into
account that each unit cell of CuNCN contains four Cu-Cu interactions
along the $a$-direction, and $a_{0},\, b_{0},\,\mathrm{and}\, c_{0}$
are the orthorombic (?) lattice constants, so that we get the estimate:
\[
K_{aa}=\frac{b_{0}c_{0}}{4a_{0}}C_{aa,aa}.
\]
Completely analogous consideration of the deformation parameter $2\rho_{c}=\delta c$,
yields the defomation tensor with single nonvanishing component $u_{cc}=\delta c/c_{0}$
(the factor 2 takes care of the fact that in the structure shown in
Fig. \ref{fig:CuNCN-ab-layer} the Cu-Cu distance in the \emph{c}
direction fits twice in the unit cell) yields: 
\[
\frac{1}{2}C_{cc,cc}a_{0}b_{0}c_{0}\left(\frac{\delta c}{c_{0}}\right)^{2}=\frac{1}{2}C_{cc,cc}a_{0}b_{0}c_{0}\left(\frac{2\rho_{c}}{c_{0}}\right)^{2}=4\times\frac{1}{2}C_{cc,cc}\frac{a_{0}b_{0}}{c_{0}}\rho_{c}^{2}=4\times\frac{1}{2}K_{cc}\rho_{c}^{2},
\]
where the factor 4 in the rightmost expresion as in the case of the
lattice direction \emph{a} takes into account the presence of the
four Cu-Cu bonds in the \emph{ab} cross section. Thus: 
\[
K_{cc}=\frac{a_{0}b_{0}}{c_{0}}C_{cc,cc}.
\]
In the general case when both lattice parameters \emph{a} and \emph{c}
change although not \emph{b} and neither of angles the deformation
tensor has two nonvanishing components $u_{aa}$ and $u_{cc}$ defined
above. The energy of one unit cell under such deformation reads
\[
a_{0}b_{0}c_{0}\left(\frac{1}{2}C_{aa,aa}\left(\frac{\delta a}{a_{0}}\right)^{2}+\frac{1}{2}C_{cc,cc}\left(\frac{\delta c}{c_{0}}\right)^{2}+C_{aa,cc}\left(\frac{\delta a}{a_{0}}\right)\left(\frac{\delta c}{c_{0}}\right)\right)
\]
Singling out the remaining off-diagonal term we write:
\[
b_{0}C_{aa,cc}\delta a\delta c=b_{0}C_{aa,cc}2\rho_{a}\rho_{c}=4\times4\times K_{ac}\rho_{a}\rho_{c}
\]
where two factors of 4 in the rightmost term take into account the
fact the four Cu-Cu bonds are extended from the given unit cell in
either direction \emph{a} or \emph{c}. Thus 
\[
K_{ac}=\frac{b_{0}}{8}C_{aa,cc}
\]

From the VASP calculations on CuNCN in various antiferromagnetic states
we have for $C_{aa,aa}$ the estimates of 195.6 or 181.3 GPa, $C_{cc,cc}$
is 159.6 GPa, and $C_{aa,cc}$ 49.5 GPa \cite{Stoffel-private}. Taking
into account the SI units relations (1 GPa = 10$^{9}$ J/m$^{3}$,
1Å=10$^{-10}$ m, $k_{B}=1.38$$\cdot$10$^{-23}$ J/K and the values
of the lattice constants of CuNCN ($a_{0}$ = 2.99, $b_{0}$ = 6.19,
$c_{0}$ = 9.41 Å, see Refs. \cite{Dronskowski,Tch099}) we obtain
the elastic constant $K_{aa}$ to be in the range of 64100 to 69000
K/Å$^{2}$ ($i.e.$ \emph{ca.} 6 to 7 eV/Å$^{2}$) pretty smaller
than the characteristic values derived for analogous constant in polyenes
\cite{MisurkinOvchinnikov}, which one, however, could expect provided
the difference between the concerned deformations of intramolecular
bonds in polyene and somewhat weaker interionic interactions in CuNCN.
Two other force constans get estimates $K_{cc}=$ 22750 K/Å$^{2}$
and $K_{ac}=$ 2775 K/Å$^{2}$. With that large difference between
the diagonal and off-diagonal force constants we can sometimes assume
the matrix $K^{-1}$ to be diagonal with elements equal to inverse
diagonal force constants $K_{aa}^{-1}$ and $K_{cc}^{-1}$.
\end{document}